\documentclass[preprint,showpacs,preprintnumbers,amsmath,amssymb]{revtex4}
\usepackage{graphicx,epsfig,dcolumn,bm,epic,eepic,float}%
\usepackage{amsmath}
\usepackage{subfigure}
\usepackage{latexsym}
\usepackage{color}
\usepackage{cancel}
\usepackage{makeidx,shortvrb,latexsym}
\usepackage{slashed}
\begin{document}
\title {On validity of different PDFs sets using the  proton $k_t$-factorization structure functions and the Gaussian $k_t$-dependence of KMR UPDFs}
\author{$Z.\; Badieian\; Baghsiyahi$ } 
\author{$M. \; Modarres$ }
\altaffiliation {Corresponding author, Email:
mmodares@ut.ac.ir, Tel:+98-21-61118645, Fax:+98-21-88004781.}
\author{$R.\; Kord\; Valeshabadi$ }
\affiliation {Department of Physics, University of $Tehran$,
1439955961, $Tehran$, Iran.} \ \
\begin{abstract}
In this work, we discuss: (i) The ratios of different parton distribution functions (PDFs), i.e., MMHT2014, CJ15, CTEQ6l1, HERAPDF15, MSTW2008, HERAPDF20 and MSHT20,  and the corresponding Kimber-Martin-Ryskin (KMR) unintegrated parton distribution functions (UPDFs) sets versus the hard scale $Q^2$, to find out the sensibility of the KMR UPDFs with respect to the input PDFs sets. It is shown that  there is not much difference between the different input-PDFs or corresponding UPDFs sets ratios. (ii) Then,  the dependence of proton $k_t$-factorization structure functions  on the different   UPDFs sets which can use the above PDFs sets as input, are presented. The results are compared with the experimental data of ZEUS, NMC and H1+ZEUS at the hard scale $Q^2 =27$ and $90$ $GeV^2$, and a reasonable agreement is found, considering different input PDFs sets.   (iii) Furthermore, by fitting  a Gaussian function, which depends on  the transverse momentum $k_t$,  to the KMR UPDFs and averaging over $x$ (the fractional parton momentum), we obtain the average transverse momentum, $<k_t^2>$, in the scale range $Q^2=2.4-1.44\times10^4$ $GeV^2$, which is in agreement with the other groups predictions, $0.25-0.44$ $GeV^2$ at $Q^2=2.4$ $GeV^2$. (iv)  Finally we explore  the average transverse momentum for which, the  results of  proton structure function with the KMR UPDFs and that of the Gaussian $k_t$-dependent, are consistent to each other. Through the above  report, at each step the parton branching (PB) UPDFs, i.e., the transverse momentum  dependent PDFs (PB TMDPDFs) are considered for comparisons with the corresponding KMR UPDFs output.
 \end{abstract}
   
\maketitle
\section{\textbf{Introduction}}
Understanding of the  proton  structure function (SF) is crucial for obtaining an accurate description of the proton-proton collisions experimental data for example at the LHC \cite{ATLAS,LHC,CMS2021}. The parton distribution functions (PDFs) \cite{ATLAS,LHC,CMS2021},  are the main elements in understanding of the structure of  hadrons and also are essential ingredients in the phenomenological calculations of high energy physics experiments. The PDFs are mostly obtained by the global analysis of hard scattering data in the collinear factorization framework. Within this framework,  the PDFs ($a(x, \mu^2)$, $a=g, q$ and $\bar{q}$) depend on $x$, the longitudinal momentum fraction of the partons with respect to   proton, and $\mu^2=Q^2$, the hard scale. In the collinear factorization it is assumed that partons have negligible transverse momentum, i.e., $k_t^2 \ll \mu^2$, and the PDFs follow the Dokshitzer–Gribov–Lipatov–Altarelli–Parisi (DGLAP) \cite{DGLAP1,DGLAP2,DGLAP3,DGLAP4} evolution equation. However, the parton transverse momentum becomes important at the high energy limit, where $x$ is small, and therefore it is  needed to have the transverse momentum dependent (TMD) PDFs (TMDPDFs), or unintegrated parton distribution functions (UPDFs) \cite{CMS2021}, which are based on the DGLAP \cite{DGLAP1,DGLAP2,DGLAP3,DGLAP4} and Balitskii-Fadin-Kuraev-Lipatov (BFKL) \cite{BFKL1,BFKL2,BFKL3} equations. 

The UPDFs ($f_a(x, k_t^2, \mu^2)$) depend on the transverse momentum $k_t$ of parton, besides having dependency on $x$ and $\mu^2$, and are historically defined for gluon which satisfies the Catani–Ciafaloni-Fiorani–Marchesini (CCFM) \cite{ccfm1,ccfm2,ccfm3,ccfm4} evolution equation. However, in order to obtain the quark UPDFs one needs to use other methods such as Kimber-Martin-Ryskin (KMR) \cite{KMR,KMR1} or Martin-Ryskin-Watt (MRW) \cite{MRW} or parton branching (PB) \cite{PB1,PB2,PBTMD} approaches.

The general  behavior and stability of the KMR and MRW models are investigated in details in the references \cite{Modarres1,Modarres2,Modarres3,Modarres4}, where these two models show promising results in describing the experimental data \cite{Modarres5,Modarres6,Modarres7,Modarres8,Modarres9}. As we mentioned above,  the  KMR and MRW formalisms are based on the DGLAP and BFKL evolution equations, as well as the Sudakov form factor,  and the collinear PDFs as their input. They  become scale ($\mu^2$) dependent only at the last evolution step. The ambiguity about the differential forms of KMR and MRW is discussed in the reference \cite{kordp}

The KMR and MRW frameworks are generally different from each other by the various imposition of angular ordering constraints and this leads to relatively small different behavior of these models in the phenomenological studies.  Also because the collinear input PDFs play an important role in these UPDFs models, therefore the investigation of various PDFs sets are vital in any phenomenological analysis. Here, in this work for investigation of their dependency to the PDFs sets, in the $k_t$-factorization framework, we only consider the KMR model where in the references \cite{Modarres2014,Modarres2019} it is shown that KMR model UPDFs   performs better than the MRW one,  in describing the deep inelastic proton structure functions, $F_2(x, Q^2)$. In contrast to the KMR and MRW models, within the PB model it is claimed that, the transverse momentum dependency comes into play from the start of the evolution ladder to obtain the UPDFs global analysis. This model also shows promising results in describing the experimental data \cite{PB_Drell,Modarres8,PBTMD}. As a result of having different methods for obtaining the UPDFs, therefore it is important to test these different approaches against the experimental data to gain a better insight of their   behaviors and  performances in various energy limits, i.e., $\mu^2$, especially in the $k_t$ factorization framework.

In addition to the aforementioned UPDFs models described above, it is also conventional to consider Gaussian distribution for the $k_t$ dependency of  UPDFs. For example in calculating the cross section of the semi-inclusive deep-inelastic-scattering (SIDIS) processes, the unintegrated fragmentation functions (UFFs)  \cite{Metz,Signori,Miguel,Anselmino1,Anselmino2,Schweitzer,T1}, and also the UPDFs with the Gaussian distribution are considered, where for the UPDFs, it is usually written as below:
\begin{equation}
	f_a(x,k_t^2,\mu^2) =f_a(x,\mu^2){e^{-{k_t^2 \over <k_t^2>}} \over \pi<k_t^2>}\label{q1}.
\end{equation}
It should be mentioned that, these distributions can also describe the data at small scales limit, and the cross section calculations with these Gaussian distributions can cover the data well \cite{T2}. Therefore, it is   interesting to investigate this distribution at larger scales and also examine the dependency of the average transverse momentum $ <k_t^2> $ by   changing the hard scale $\mu^2$ and the $x$ value for the UPDFs. 

The purpose of the present work is to   investigate the sensitivity of KMR UPDFs to the various input sets of PDFs. For this reason, after obtaining the ratio of different sets of PDFs and UPDFs, we calculate the proton $k_t$ factorization structure function for these UPDFs sets to find out which PDFs (UPDFs) sets give the closer results regarding   the available experimental data, i.e., NMC \citep{NMC}, ZEUS \citep{ZEUS} and H1+ZEUS \citep{H1+ZEUS}. The SF with the PB TMDPDFs \cite{PB1,PB2,PB_Drell,PBTMD}  are also calculated and the PB TMDPDFs ratios to the KMR UPDFs are   discussed, as well.

Therefore, the structure of the paper is arranged as follows: In the subsections \ref{subsection 2-1} and \ref{subsection 2-2},  we present  the KMR UPDFs method and the calculation of PB TMDPDFs, as well as   the  proton $k_t$-factorization structure function, respectively.  The ratio of UPDFs using various PDFs sets for different values of $x$ and transverse momentum $k_t$,   and  the results of fitting the Gaussian function to the  KMR UPDFs     are given in the subsections \ref{subsection3.1} and \ref{subsection3.2}, respectively.  We also plot the variation of average transverse momentum versus the  $ x $ values for the two different energy scales. The results  of  proton structure functions  for different UPDFs models, i.e., the KMR, PB and Gaussian transverse momentum dependent UPDFs models are given in the subsection \ref{subsection3.3}. Finally, in the section IV our conclusions will be presented. 
\section{The general formalisms for the KMR UPDFs and the PB TMDPDFs calculation and the proton $k_t$-factorization structure functions    }
 \subsection{The UPDFs of KMR method and the PB TMDPDFs} \label{subsection 2-1}
The KMR UPDFs, $f_a(x, k_t^2, \mu^2)$, can be calculated through the KMR \citep{KMR,KMR1} procedure, based on   the DGLAP and BFKL approaches. The  inputs are the single-scale  PDFs, i.e.,  $a(x,\mu^2)=xq(x,\mu^2),  x\bar{q}(x,\mu^2), xg(x,\mu^2)$, and the leading order (LO) splitting functions $P_{aa^\prime} (x)$. The modified   DGLAP equations is usually written as \citep{KMR,KMR1},
\begin{equation}
\frac{\partial a(x,\mu^2)}{\partial \ln \mu^2}=\frac{\alpha_s}{2\pi} \left[\int_{x}^{1-\Delta}dz P_{aa^\prime}(z) a^\prime(\frac{x}{z},\mu^2)-a(x,\mu^2)\sum_{a^\prime}\int_{0}^{1-\Delta} P_{a^\prime a}(z^\prime)dz^\prime \right],\label{2}
\end{equation}
where the cut-off $\Delta$,    prevents the $z = 1$ singularities in $P_{aa^\prime}(x)$  arises from the soft gluon emission (note that for $g \rightarrow gg$ splitting, an extra factor $z^\prime$ should be inserted in the front of $P_{aa^\prime}$ in the last integral of equation (\ref{2})).  The cutoff $\Delta$ can be obtained from the   well known angular ordering condition \citep{ccfm1,ccfm2,ccfm3,QCD,Kimber}, in the
last  evolution step, so,
\begin{equation}
\Theta(\theta-\theta^{\prime}) \Rightarrow \mu>\frac{zk_t}{1-z},
\end{equation}
which leads to an upper limit on the integration variable $z$, i.e.,
\begin{equation}
z_{max}=\frac{\mu}{\mu+k_t} \Rightarrow \Delta =1-z_{max} =\frac{k_t}{\mu+k_t}
\end{equation}
The re-summation of all virtual contributions, i.e., the second term of   the equation (\ref{2}),   leads to the survival probability  Sudakov form factor as,
\begin{equation}
T_a(k_t,\mu) = exp \left( - \int_{k_t^2}^{\mu^2} {\alpha_S(k^{\prime2}_t)
\over 2\pi}\frac{dk^{\prime2}_t}{k^{\prime2}_t}  \sum_{a^\prime} \int^{1-\Delta}_{0} dz^\prime
P_{a^\prime a}(z^\prime) \right).
\end{equation}
Finally  the explicit form of the  KMR UPDFs can be written  as follows:
\\ (i) for the  quarks and the anti-quarks ($q^\prime=q,\bar{q}$),
\begin{equation}\label{fq}
f_{q^\prime}(x, k_t^2, \mu^2)= T_{q^\prime}(k_t, \mu){\alpha_S(k_t^2) \over 2\pi}  \int^{1-\Delta}_{x} dz \left[ P_{q^\prime q^\prime}(z){x \over z} q^\prime\left( {x \over z}, k_t^2 \right)+ P_{q^\prime g}(z){x \over z} g\left( {x \over z}, k_t^2 \right)\right],
\end{equation}
\\ (ii) for the gluon:
\begin{equation}\label{fg}
f_g(x,k_t^2,\mu^2)= T_g(k_t,\mu){\alpha_S(k_t^2) \over 2\pi}  \int^{1-\Delta}_{x} dz \left[\sum_{q^\prime=q,\bar{q} }P_{gq^\prime}(z){x \over z} q^\prime\left( {x \over z}, k_t^2 \right)+ P_{gg}(z){x \over z} g\left( {x \over z}, k_t^2 \right)\right].
\end{equation}
The dependence of KMR UPDFs, $f_a(x, k_t^2 , \mu^2)$, on the two scales $k_t^2$ and $\mu^2$, are imposed on the last step  evolution and the angular ordering is considered.   The  KMR formalism allows us to extend smoothly, the $k_t$ dependency of UPDFs into the region where $k_t > \mu$.

For the PB TMDPDFs \cite{PB_Drell,PBTMD} we use the TMDLIB given in the reference \cite{PBLINK}, which can generate the PB TMDPDFs at various $x$ values, the hard scales, $\mu^2$, and the transverse momentum, $k_t$. One should note that the generated PB TMDPDFs through the TMDLIB  are divided by $k_t^2$ with respect to the  KMR UPDFs explained above.  
 \subsection{The proton  $k_t$-factorization Structure Function }\label{subsection 2-2}
In the deep-inelastic scattering of electron-proton, the gluon can only interact with the virtual photon through the intermediate quark or anti-quark by considering   the electromagnetic interaction. Therefore  three LO diagrams contribute to the calculation of   structure function $F_2(x,Q^2)$. The two of these diagrams are related to $g\rightarrow q\bar{q}$ which are shown as the box and the cross-box in the references \citep{KMR,KMR1,Modarres2014,Modarres2019}.

For calculation of the gluon contribution to the proton  SF generated from the box and cross-box diagrams (using an angular average value   $<\phi> =\frac{\pi}{4}$ \cite{KMR,KMR1}), we have: 
\begin{equation}\begin{aligned}\label{fg-1}
F_T^{gluon\rightarrow quark}(x,Q^2)&=\Sigma_{q=u,d,s,c} e_q^2 \frac{Q^2}{4\pi} \int_{k_0^2}^{k_{max}^2} \frac{dk_t^2}{k_t^4} \int_{0}^{1}d\beta \int_{k_0^2}^{k_{max}^2}d\kappa_t^2\alpha_s(\mu^2)\\
&f_g(\frac{x}{\bar{z}},k_t^2,\mu^2)\Theta(1-\frac{x}{\bar{z}})\left(\left[\beta^2+(1-\beta)^2\right](J_1+J_2-J_3)+m_q^2(J_4+J_5-J_6)\right),
\end{aligned}
\end{equation}
and
\begin{equation}\begin{aligned}\label{fg-2}
F_L^{gluon\rightarrow quark}(x,Q^2)&=\Sigma_{q=u,d,s,c} e_q^2 \frac{Q^2}{4\pi}\int_{k_0^2}^{k_{max}^2} \frac{dk_t^2}{k_t^4} \int_{0}^{1}d\beta \int_{k_0^2}^{k_{max}^2}d\kappa_t^2\alpha_s(\mu^2)\\
&f_g(\frac{x}{\bar{z}},k_t^2,\mu^2)\Theta(1-\frac{x}{\bar{z}})\left(4Q^2\beta^2(1-\beta)^2)(J_4+J_5-J_6)\right),
\end{aligned}
\end{equation}
where $k_{max}=4Q$,  $J$s are defined in the appendix \ref{a} and,
\begin{equation}
{1 \over \bar{z}} = 1+ \frac{\kappa_t^2+m_q^2}{(1-\beta)Q^2}+\frac{\kappa_t^2+k_t^2-2k_t\kappa_t\cos<\phi>+m_q^2}{\beta Q^2}.
\end{equation}
For the SF calculation  of the  non-perturbative part ($k_t<k_0$, $k_0=1$ $GeV$), we use:
\begin{equation}\label{fg-3}
\int_{0}^{k_0^2} \frac{d k_t^2}{k_t^2}f_g(x,k_t^2,\mu^2)[\frac{Remainder}{k_t^2}] =xg(x,k_0^2)T_g(k_0,\mu)\left[ \right]_{k_t =a},
\end{equation}
where $a$ is a number between $0$ and $k_0$  .
$\beta$, similar to $x$, is a fraction of the photon momentum which is carried by the intermediate quark and $z$ is the fraction of  gluon momentum that is transferred to the quark. If $x$ is the fraction of the proton momentum carried by the quark, then, the fraction of the proton momentum carried by gluon is equal to $\frac{x}{z}$. $k_t$ is the parent gluon transverse momentum and $\kappa_t$ is transverse momentum of daughter quark. We  ignore the mass of light quarks and take the mass of  charm quark to be  equal to $ m_c =1.4 $ GeV  and set,
\begin{equation}
 \mu^2 = k_t^2+\kappa_t^2+m_q^2.
 \end{equation}
The gluon contribution to the proton SF is equal to sum of the equations (\ref{fg-1}), (\ref{fg-2}) and (\ref{fg-3}).

For the SF of the process that the quark  radiates gluon before interaction with the  photon, the perturbative part of the  quark  contribution   is:
\begin{equation}\begin{aligned}\label{fq-1}
F_2^{quarrk \rightarrow quark}(x,Q^2)= \Sigma_{q =u,d,s,c} e_q^2 &\int_{k_0^2}^{Q^2} \frac{d\kappa_t^2}{\kappa_t^2}\frac{\alpha_s(\kappa_t^2)}{2\pi} \int_{k_0^2}^{\kappa_t^2}\frac{dk_t^2}{k_t^2} \int_{x}^{Q/(Q+k_t)} dz \\
&\left[f_q(\frac{x}{z},k_t^2,Q^2)+f_{\bar{q}}(\frac{x}{z},k_t^2,Q^2)\right]p_{qq}(z).
\end{aligned}
\end{equation}
While,  for the    non-perturbative region, the quark part of SF   is related to the state in which a quark with a distribution function $xq (x,k_0 ^ 2)$ enters the evolution chain from the non-perturbative region and interacts with the virtual photon without real radiation, i.e.,
\begin{equation}\label{fq-2}
F_2^{quark \rightarrow quark}(x,Q^2)= \Sigma_{q=u,d,s,c} e_q^2 \left[xq(x,k_0^2)+x\bar{q}(x,k_0^2)\right]T_q(k_0,Q).
\end{equation}
Then the quark contribution to the SF is  sum of  the   equations (\ref{fq-1}) and   (\ref{fq-2}).

Finally, the total proton SF ($F_2(x,Q^2)$) becomes equal to sum of the quarks and gluon contributions, i.e., the equations (\ref{fg-1}), (\ref{fg-2}),  (\ref{fg-3}), (\ref{fq-1}) and   (\ref{fq-2}). 
\section{\textbf{Results and discussions}}
\subsection{The PDFs of different groups data fitting sets ratios and their corresponding KMR UPDFs ratios }\label{subsection3.1}
\begin{figure}
\begin{center}
\subfigure[ ]{
\includegraphics[height =6.1cm]{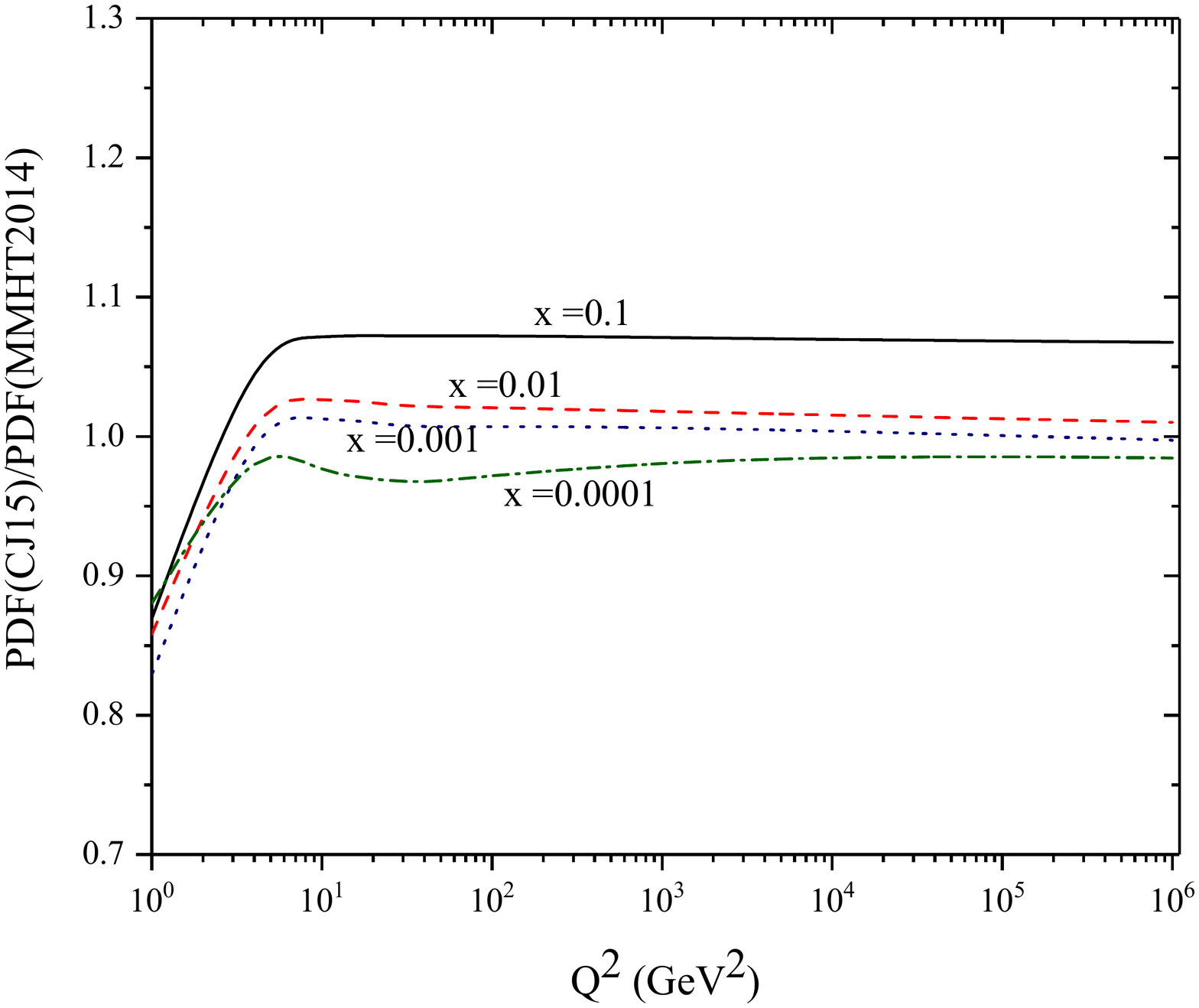}
}
\subfigure[ ]{
\includegraphics[height =6.1cm]{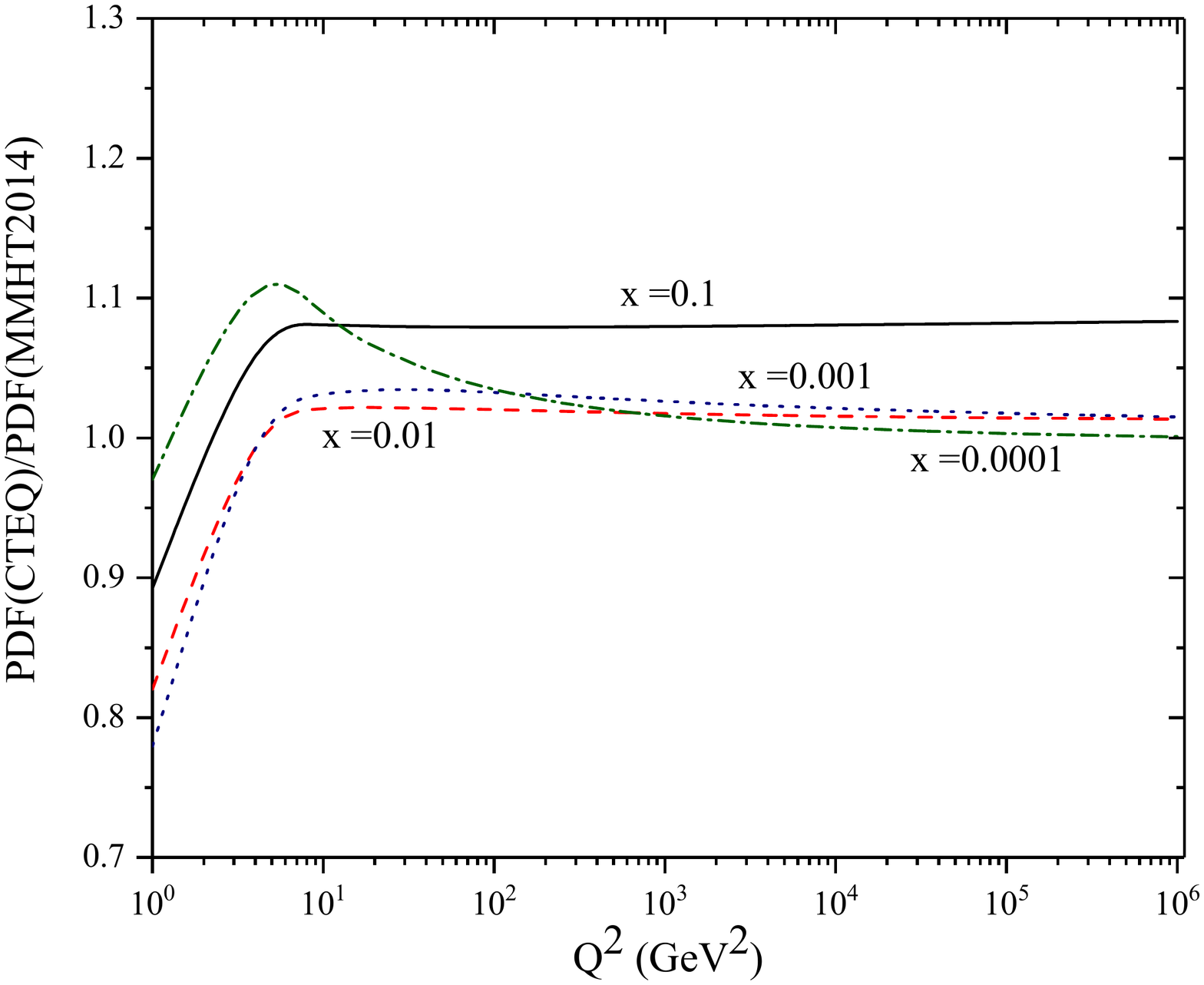}
}
\subfigure[ ]{
\includegraphics[height =6.1cm]{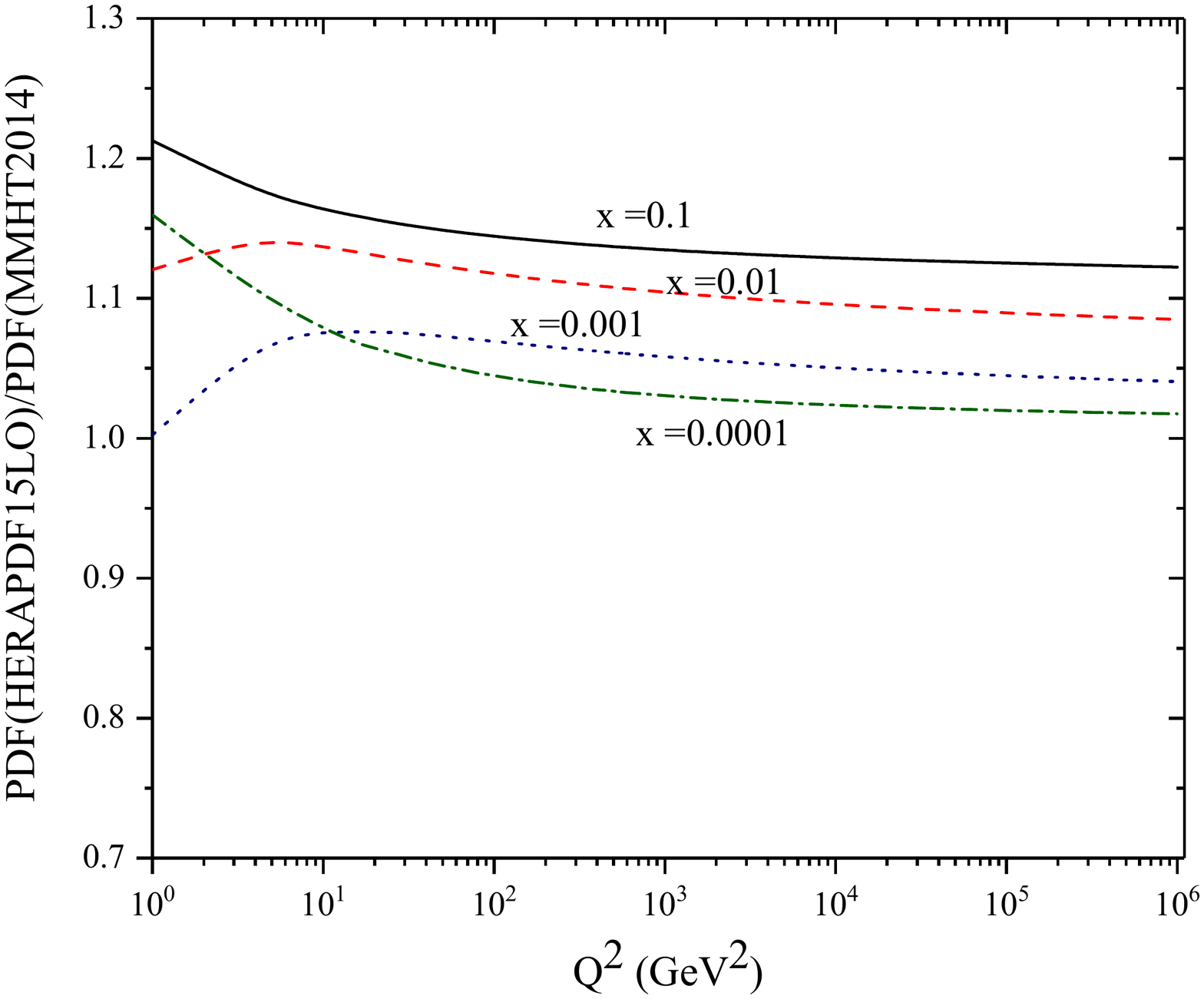}
}
\subfigure[ ]{
\includegraphics[height =6.1cm]{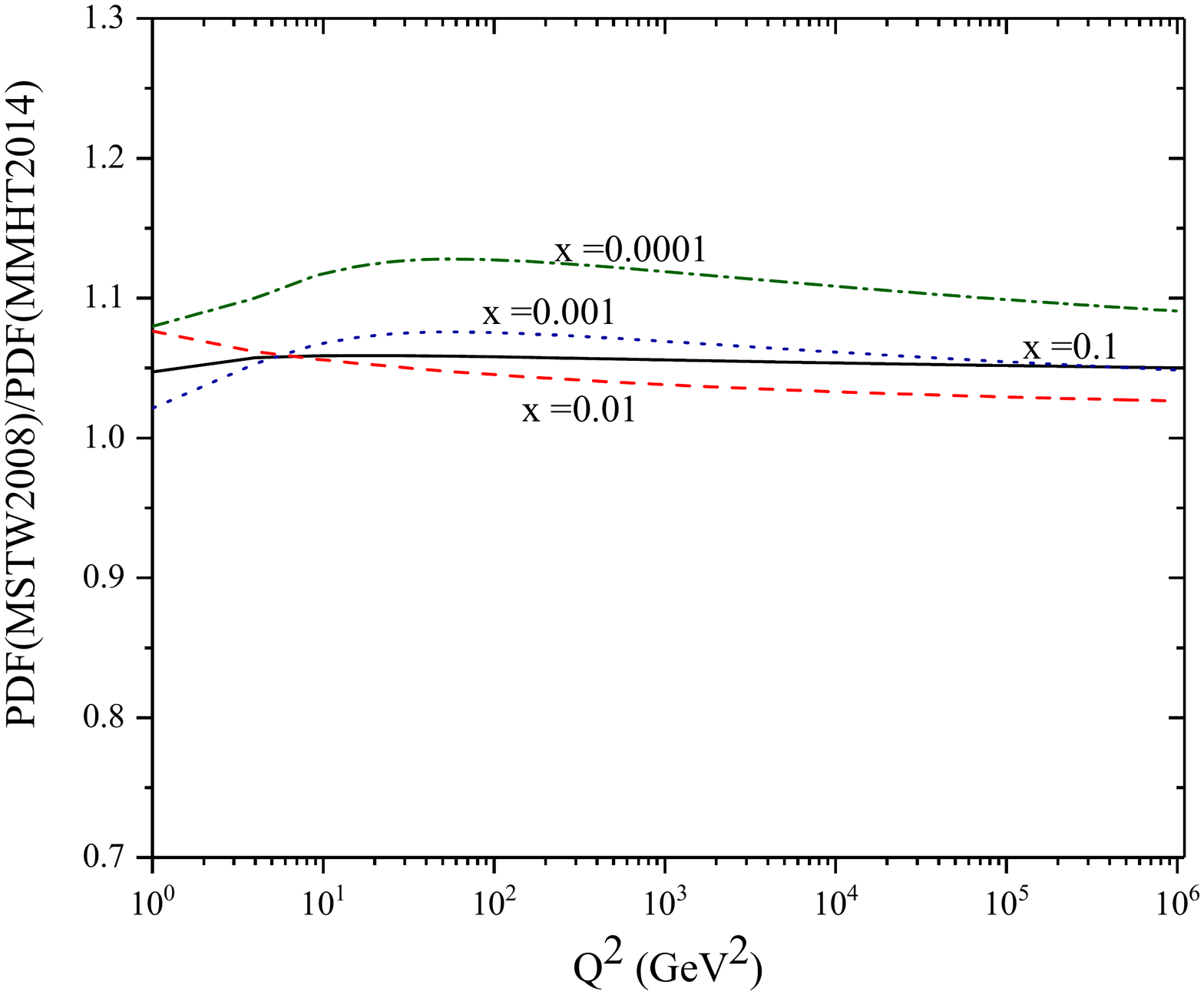}
}
\subfigure[ ]{
\includegraphics[height =6.1cm]{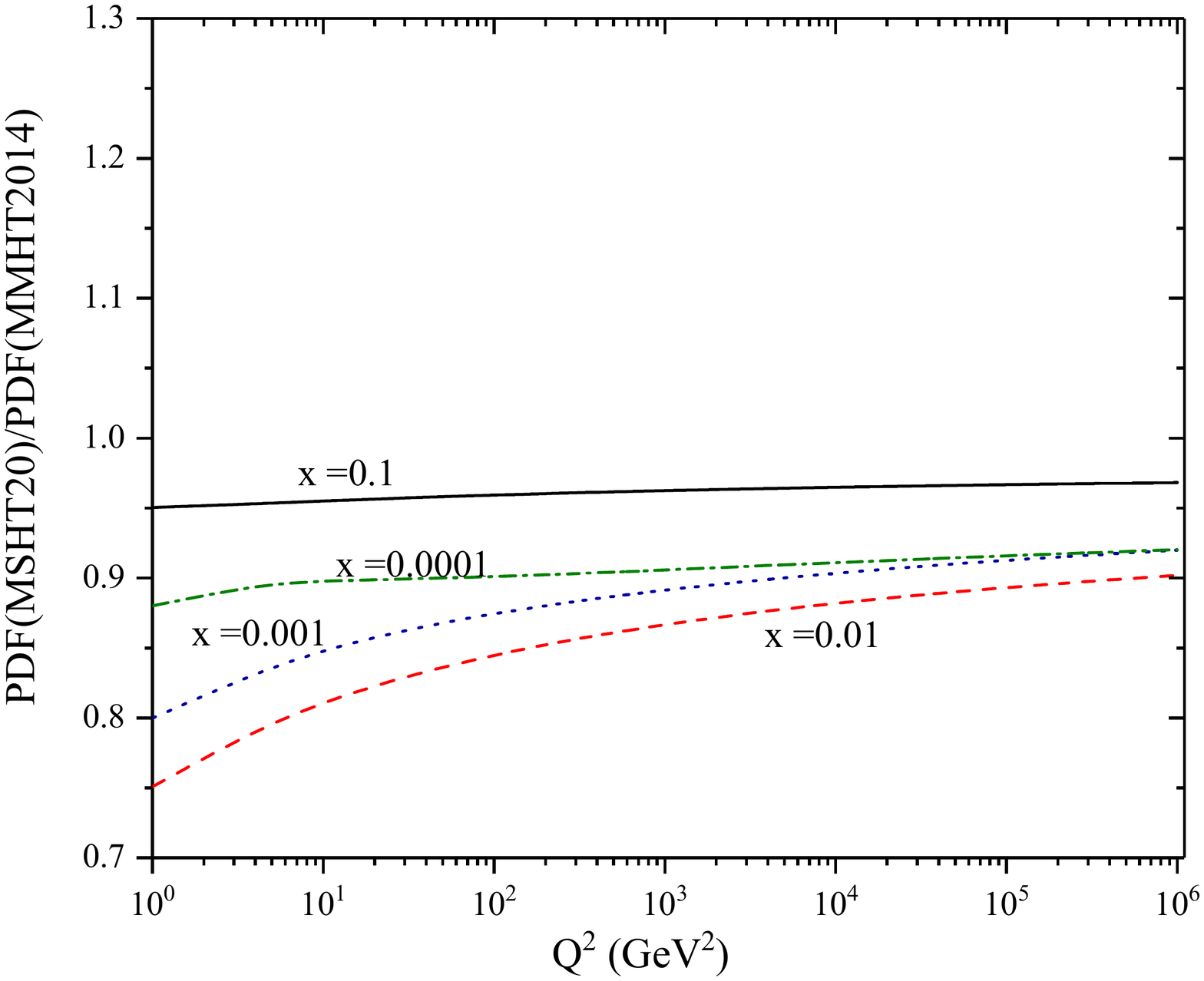}
}
\subfigure[ ]{
\includegraphics[height =6.1cm]{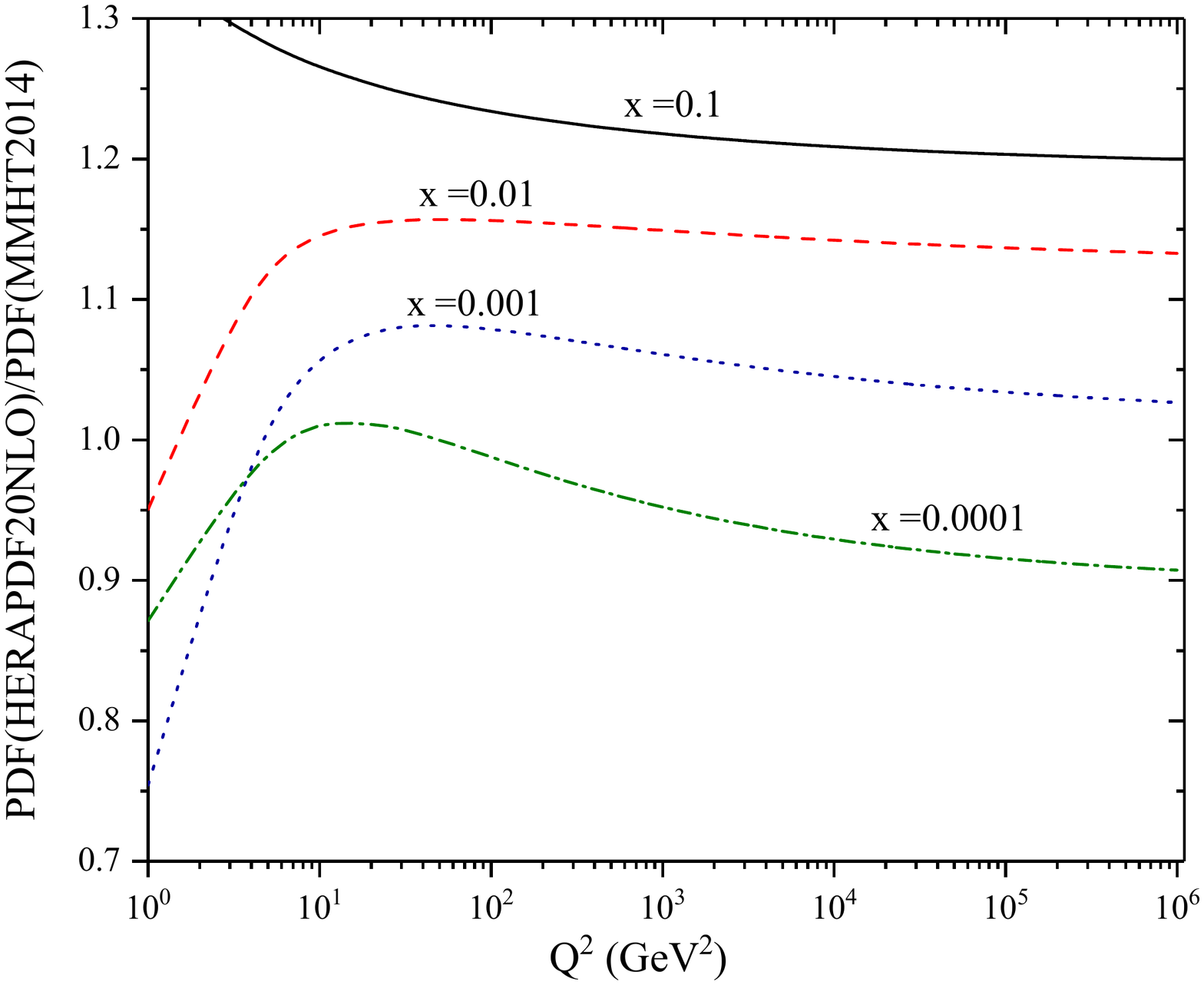}
}
\caption{The different up-quark PDFs ratios for various $x$ values given in the  panels (a), (b), (c), (d), (e) and (f).  
}\label{pdf}
\end{center}
\end{figure}

In this section the  different groups data fittings PDFs sets such as:   MMHT2014 \citep{MMHT}, 
CJ15 \citep{CJ}, CTEQ6l1 \citep{cteq}, HERAPDF15 \citep{hera}, MSTW2008 \citep{mstw}, MSHT20 \citep{msht} and HERAPDF20 \cite{heranlo}, at different $x$ values, which are generated through the   LHAPDF6 library \citep{LHAPDF} are investigated by calculating their up-quark PDFs  ratios with respect to the  MMHT2014  PDFs sets, see the figure \ref{pdf}. We chose MMHT2014 PDFs set because in our previous works it fits  reasonably the SIDIS data \cite{Modarres2019}. There is a good agreement between the different groups PDFs sets especially above $10$ $GeV^2$. Only the HERAPDF20 PDFs set shows some deviations in the whole $Q^2$ region. 

\begin{figure}
\begin{center}
\subfigure[ ]{
\includegraphics[height =6.1cm]{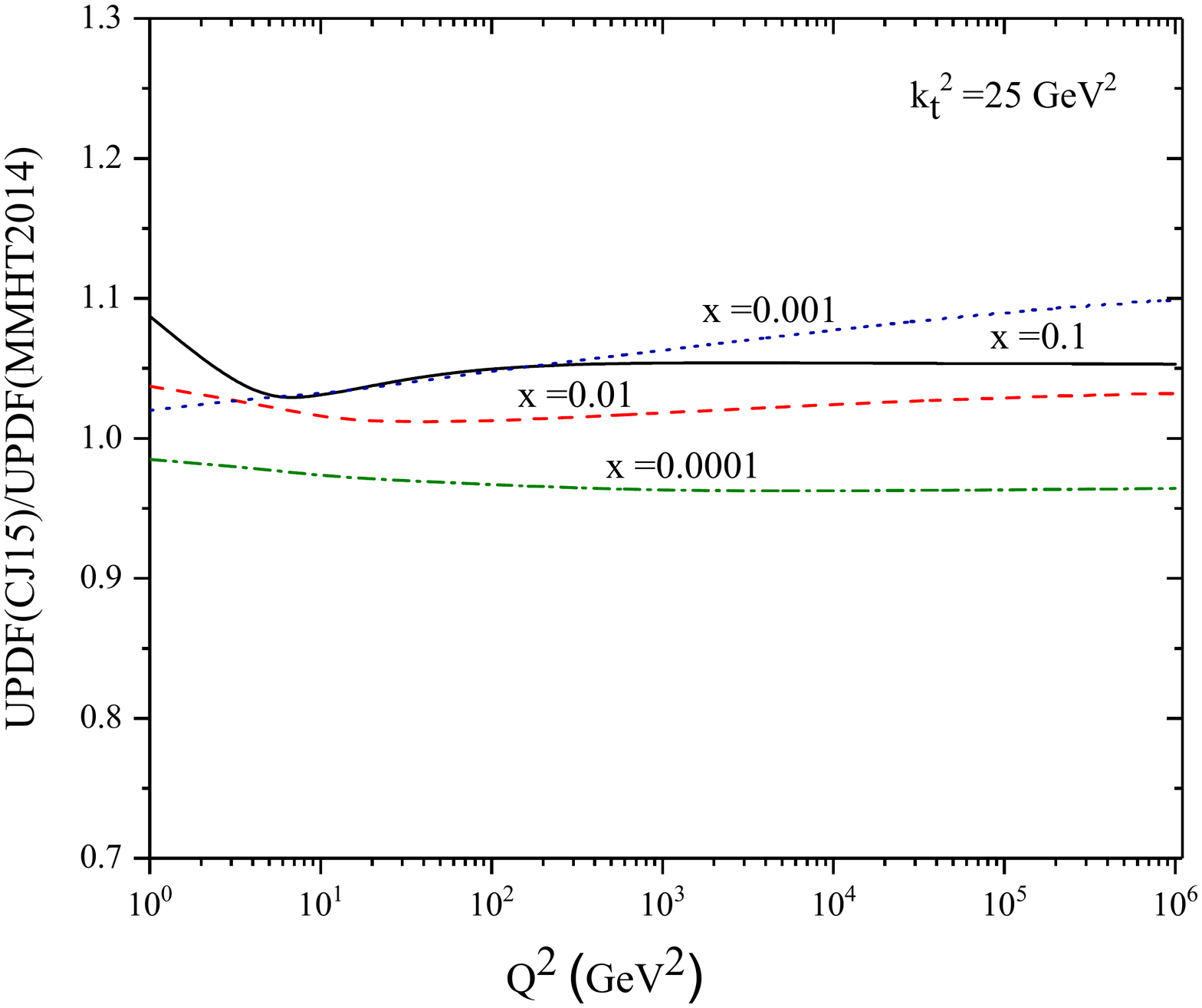}
}
\subfigure[ ]{
\includegraphics[height =6.1cm]{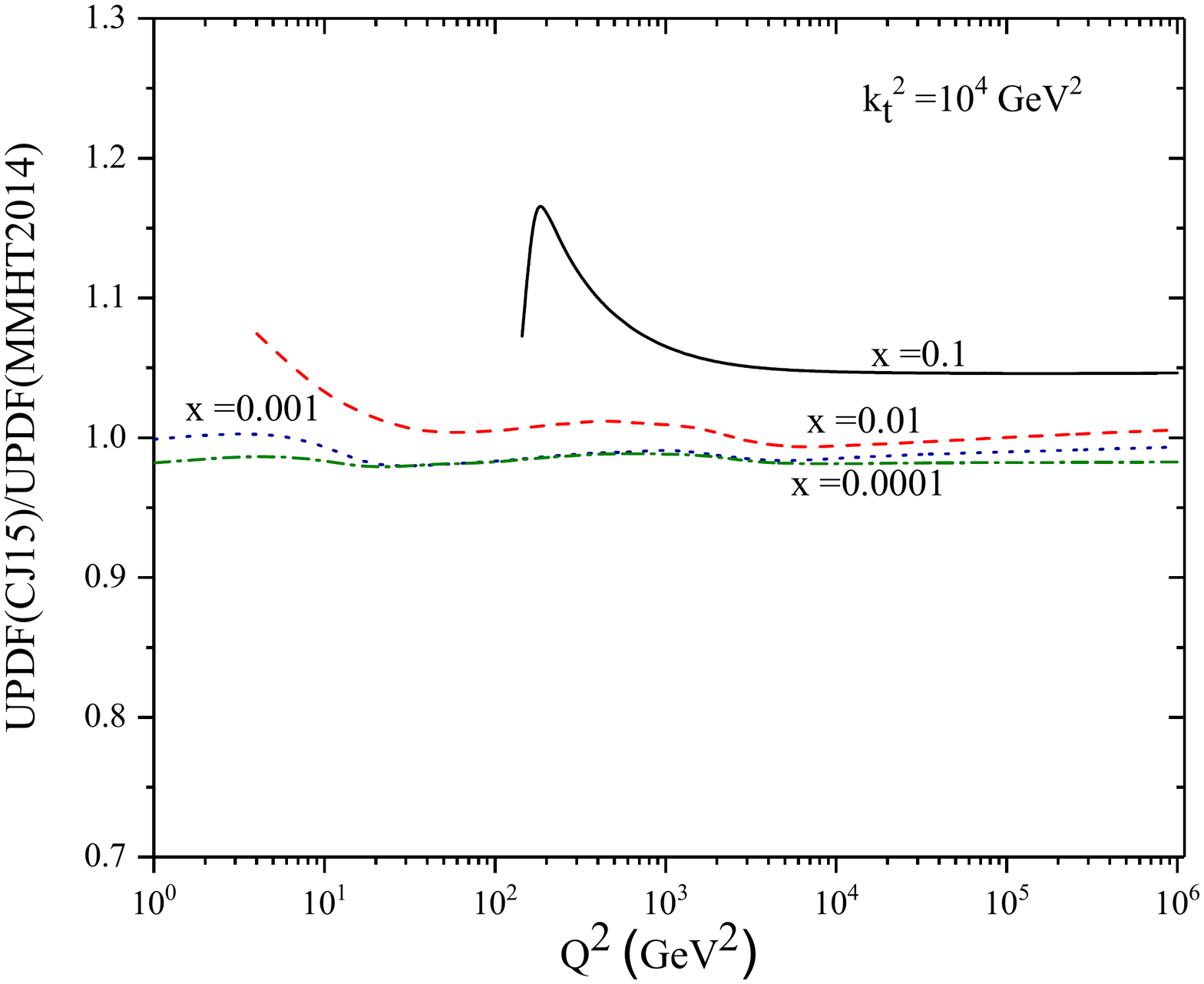}
}
\subfigure[ ]{
\includegraphics[height =6.1cm]{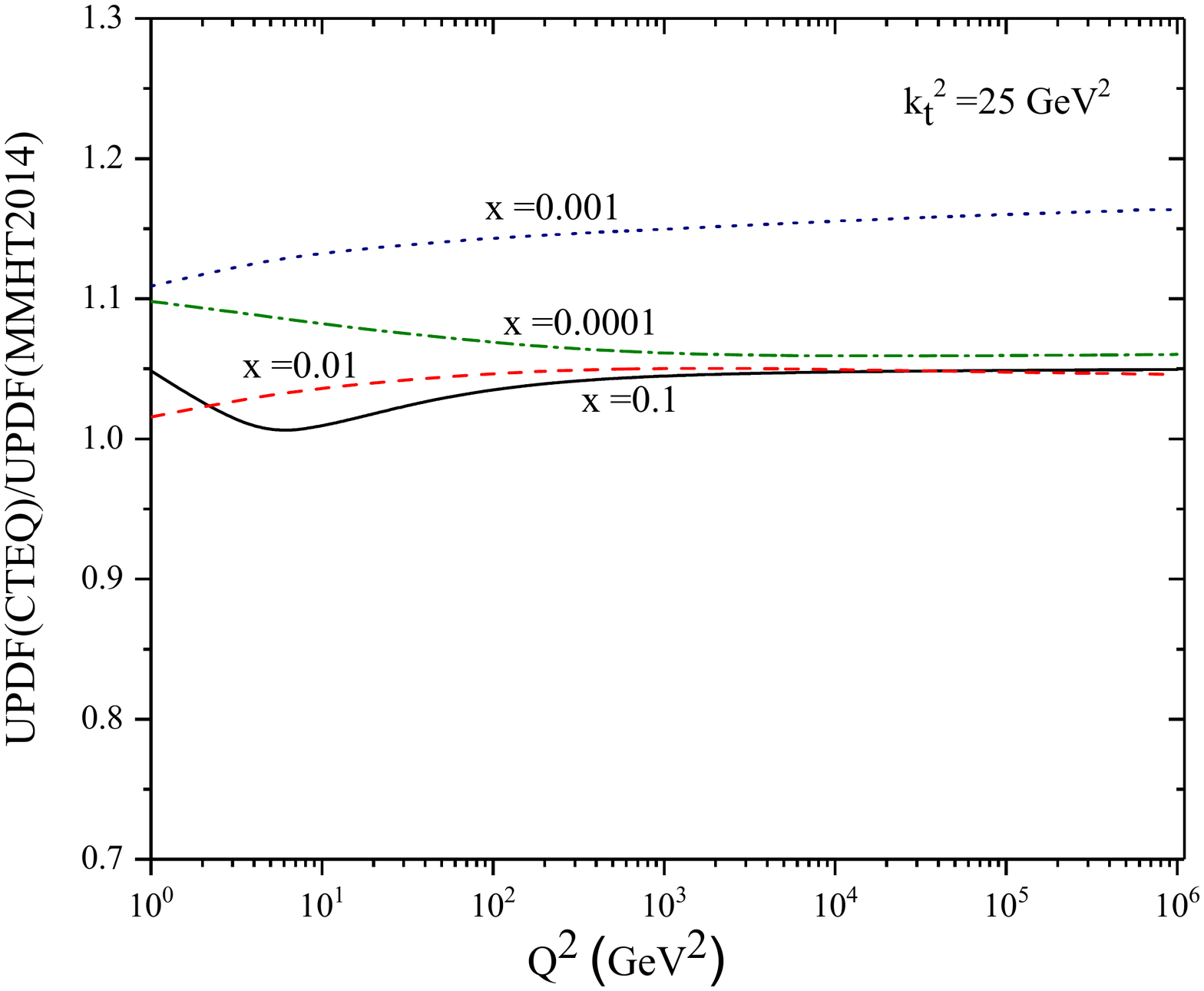}
}
\subfigure[ ]{
\includegraphics[height =6.1cm]{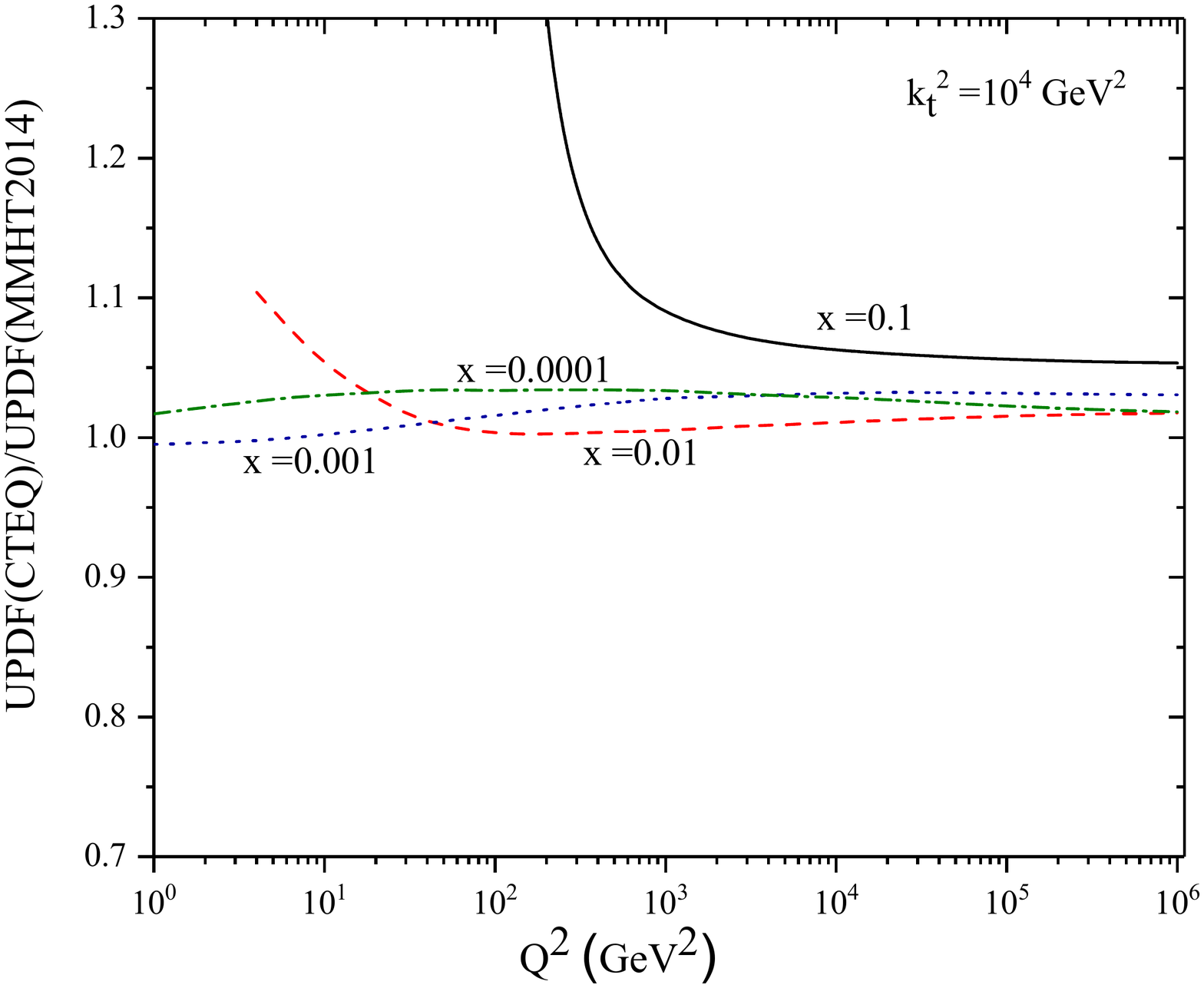}
}
\subfigure[ ]{
\includegraphics[height =6.1cm]{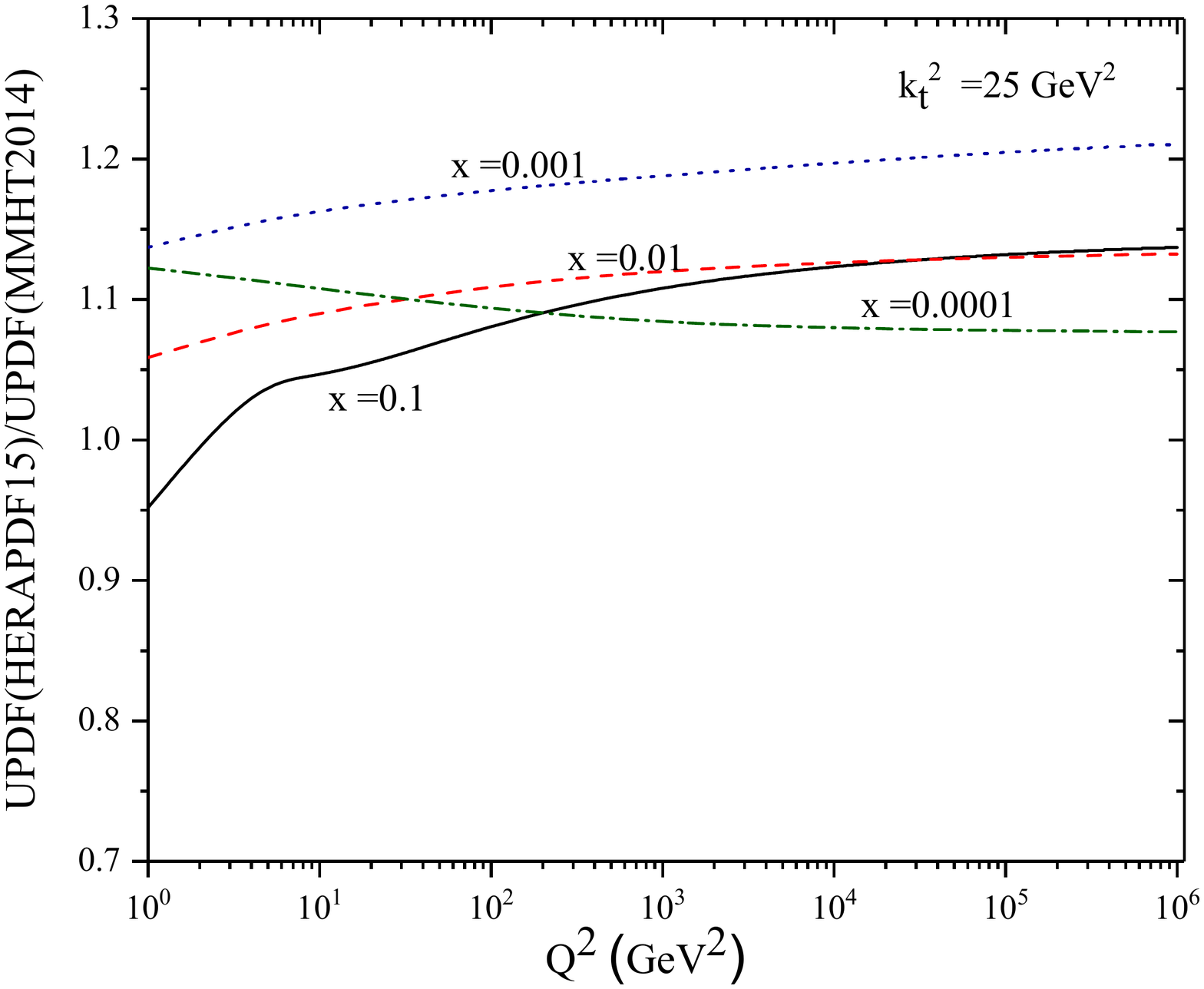}
}
\subfigure[ ]{
\includegraphics[height =6.1cm]{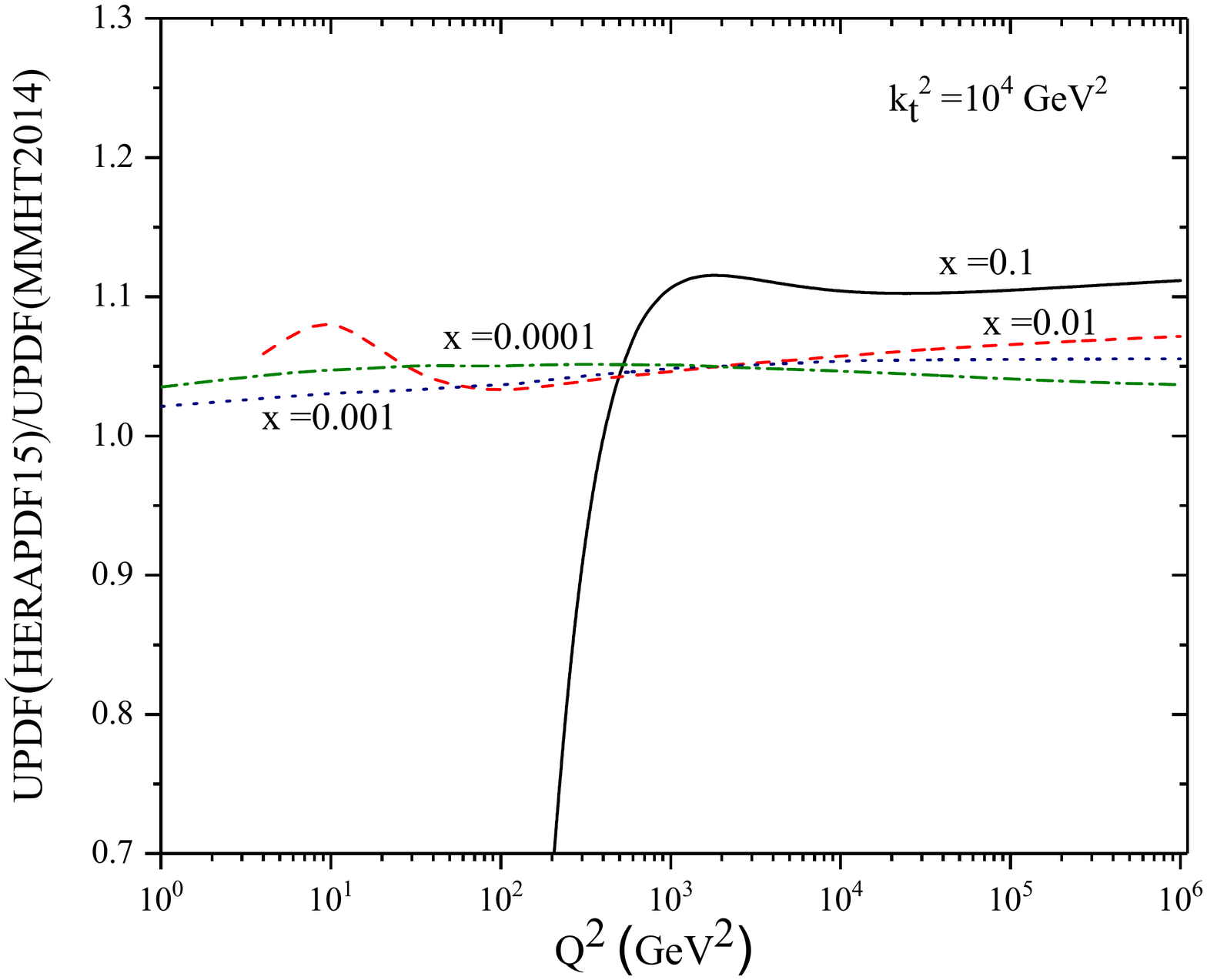}
}
\caption{The different up quark UPDFs ratios of various groups PDFs sets with respect to MMHT2014 one, at the two $k_t^2$ and   four $x$ values: The panels (a), (b), (c), (d), (e) and (f) .
}\label{updf0}
\end{center}
\end{figure}

\begin{figure}
\begin{center}
 \subfigure[ ]{
\includegraphics[height =6.1cm]{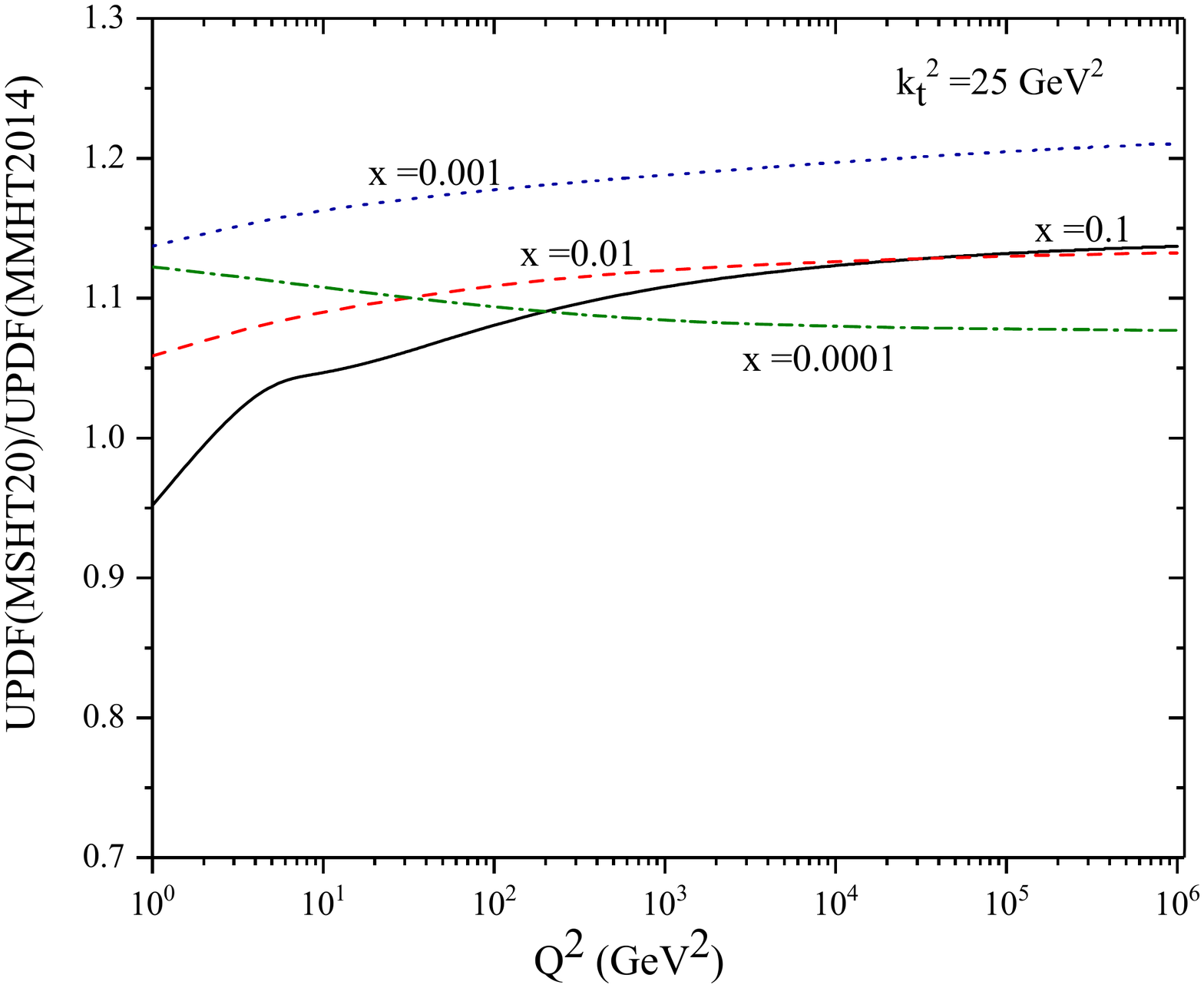}
}
\subfigure[ ]{
\includegraphics[height =6.1cm]{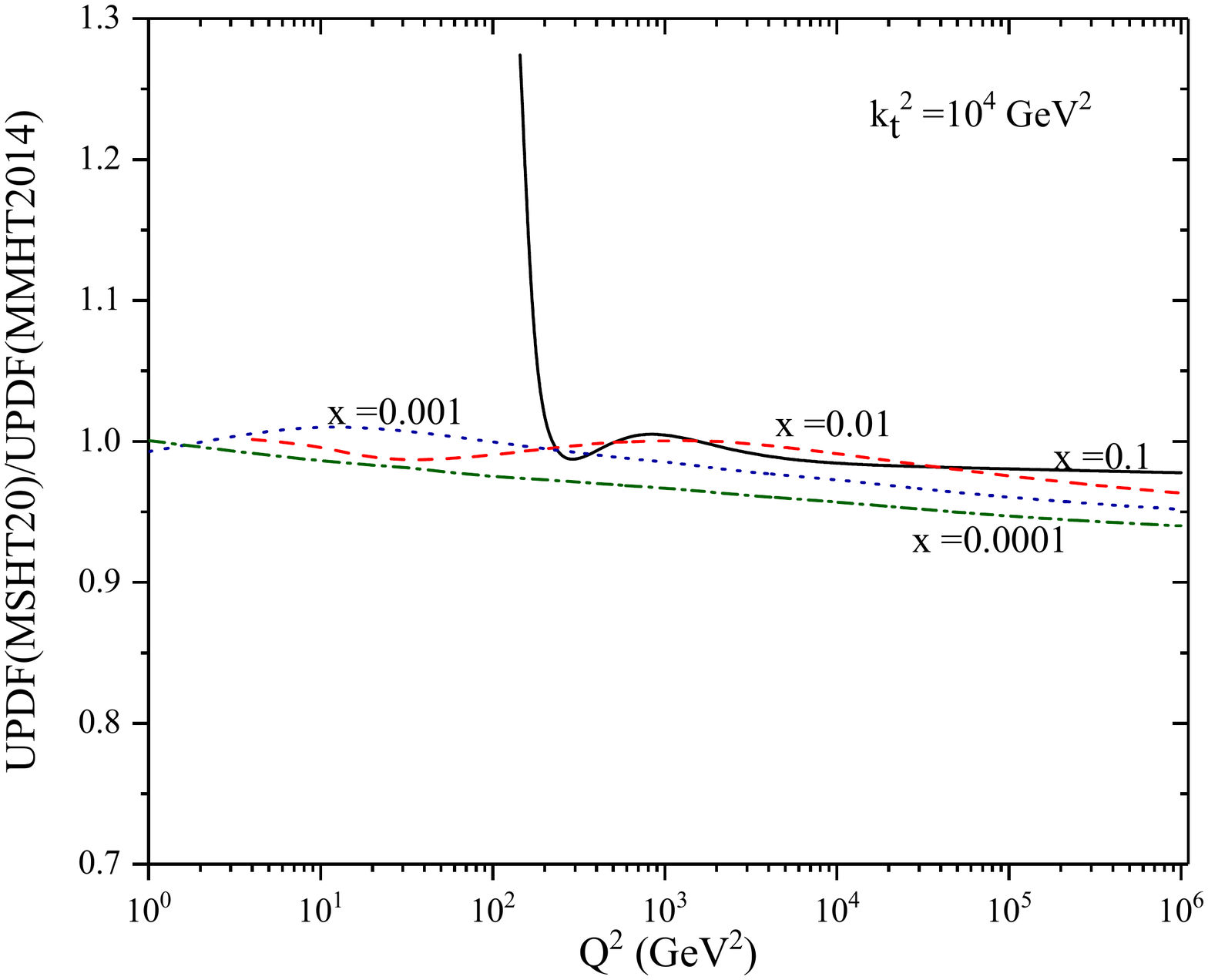}
}
\subfigure[ ]{
\includegraphics[height =6.1cm]{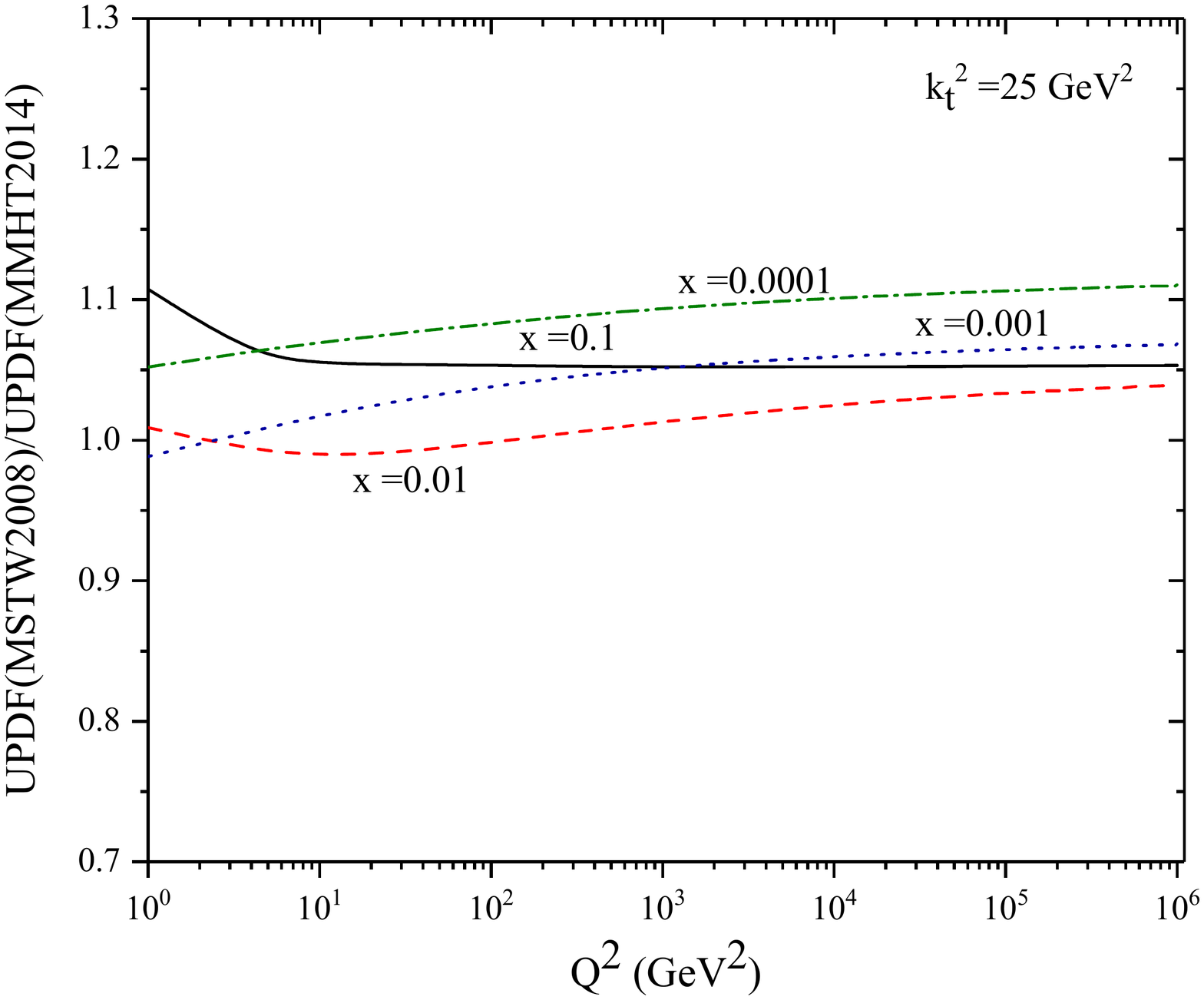}
}
\subfigure[ ]{
\includegraphics[height =6.1cm]{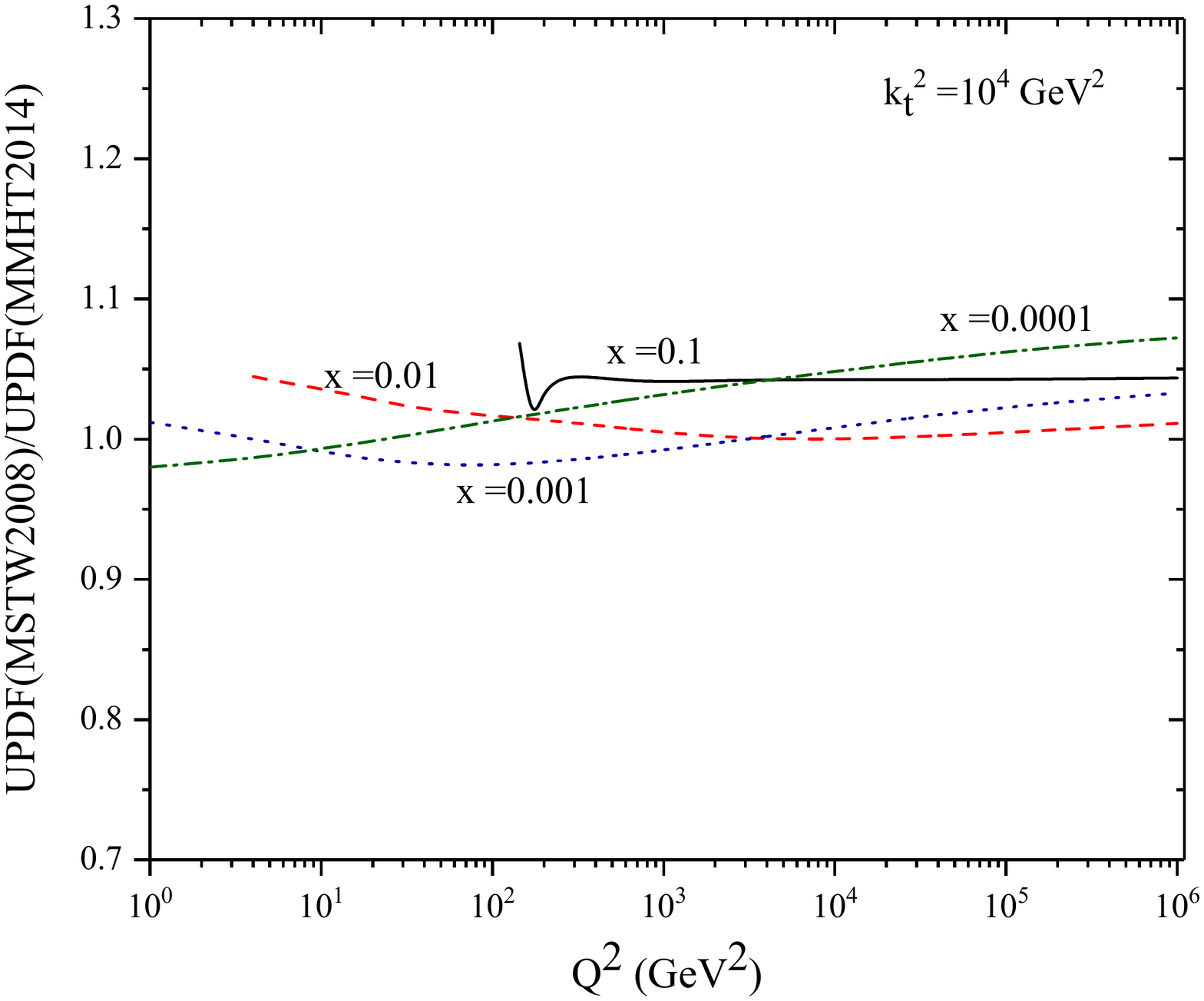}
}
\subfigure[ ]{
\includegraphics[height =6.1cm]{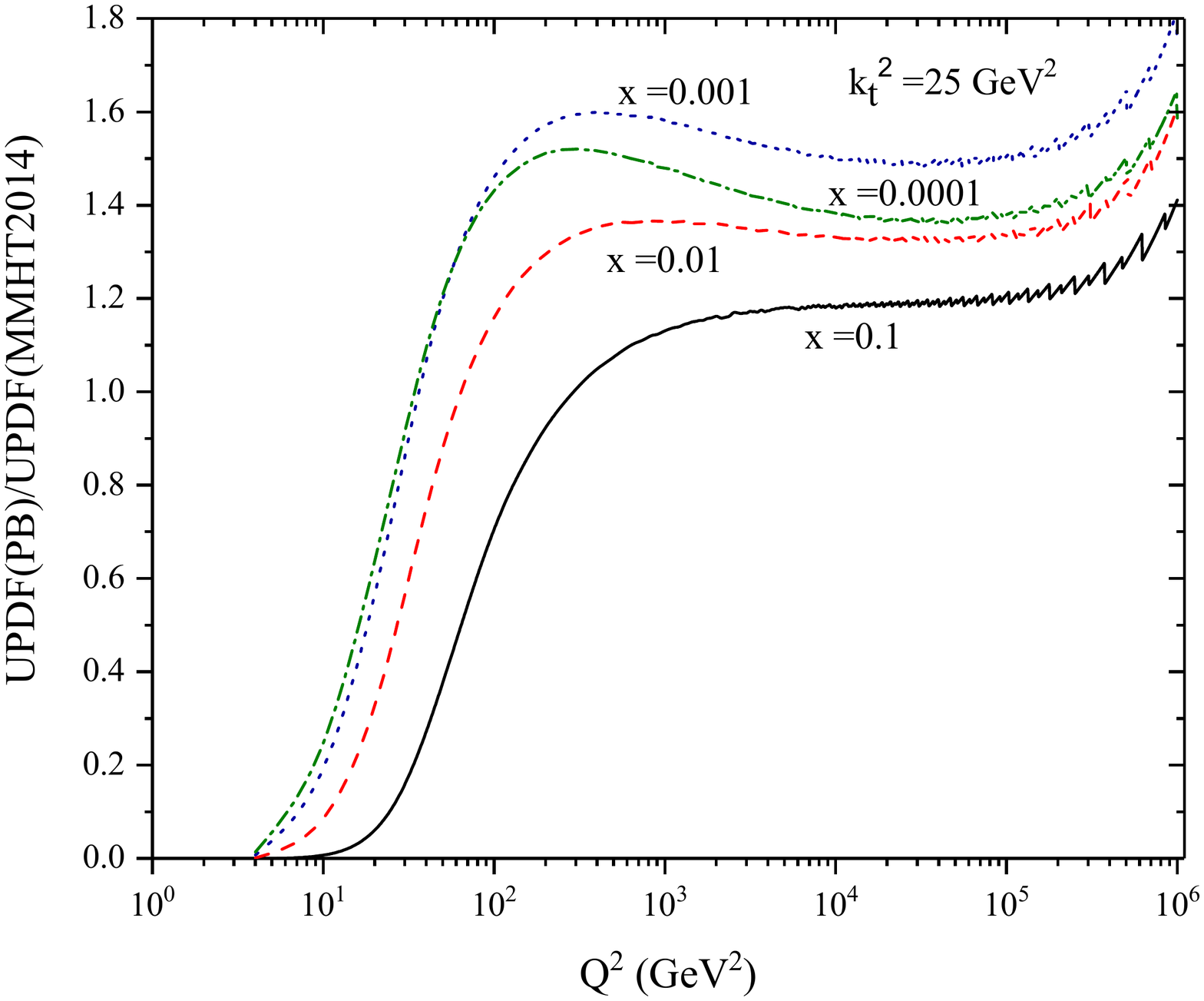}
}
\subfigure[ ]{
\includegraphics[height =6.1cm]{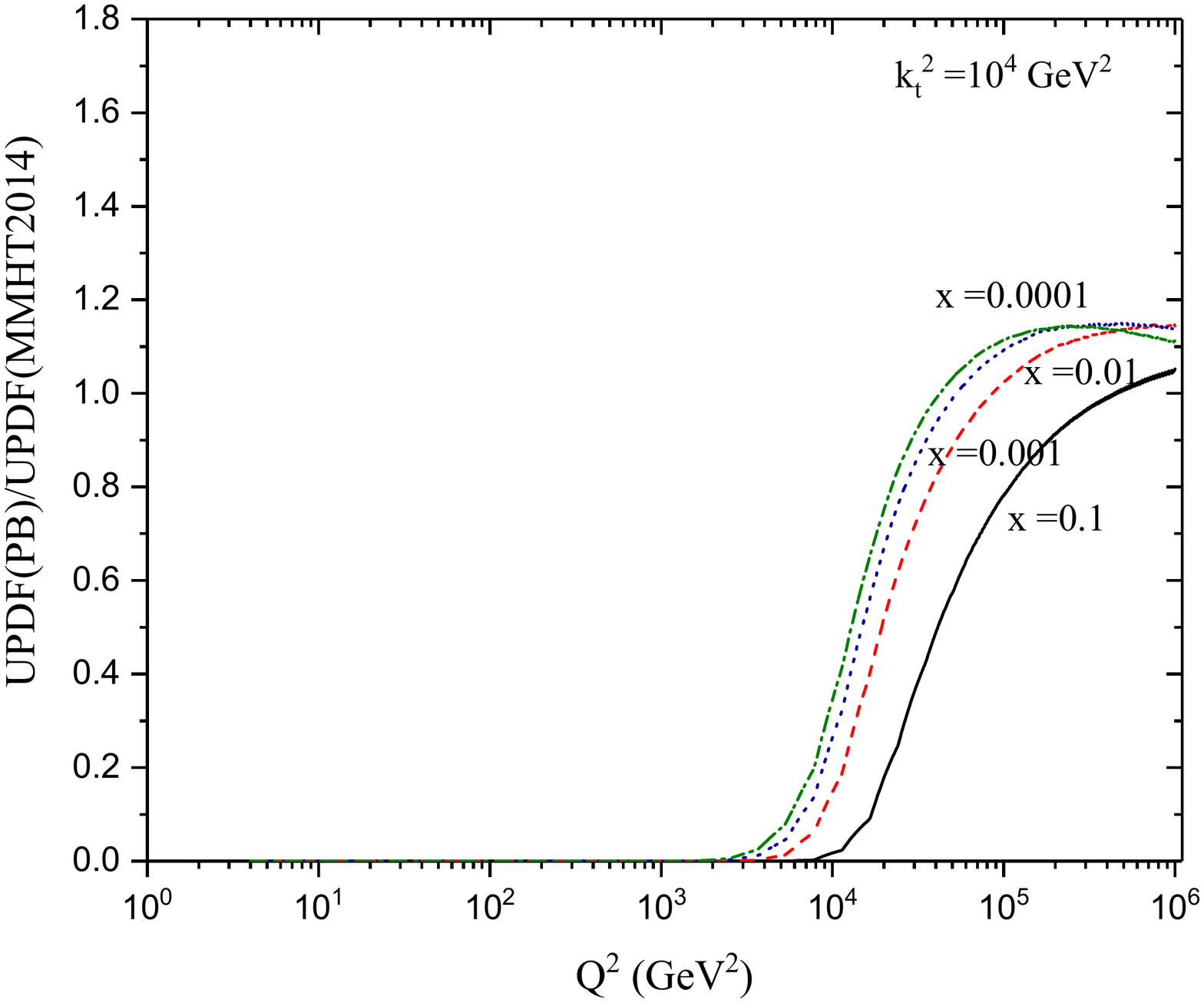}
}
\caption{The different up quark UPDFs ratios of various groups PDFs sets with respect to MMHT2014 one, at the two $k_t^2$ and  the four $x$ values: The panels (a), (b), (c), (d), (e) and (f).
}\label{updf1}
\end{center}
\end{figure}

In order to see the effect of KMR UPDFs (by using the KMR method presented in the subsection \ref{subsection 2-1}) with various inputs PDFs sets, in the figures \ref{updf0} and \ref{updf1} the  obtained up-quark UPDFs ratios  with respect to the KMR UPDF(MMHT2014)     for    the longitudinal momentum fraction ${x =10^{-1},10^{-2},10^{-3},10^{-4}}$ and the transverse momentum $k_t^2 =25$ and $10^4$  $GeV^2$ are plotted    in terms of the hard scale $Q^2$. 
\begin{figure}
\begin{center}
{
\includegraphics[height =6.1cm]{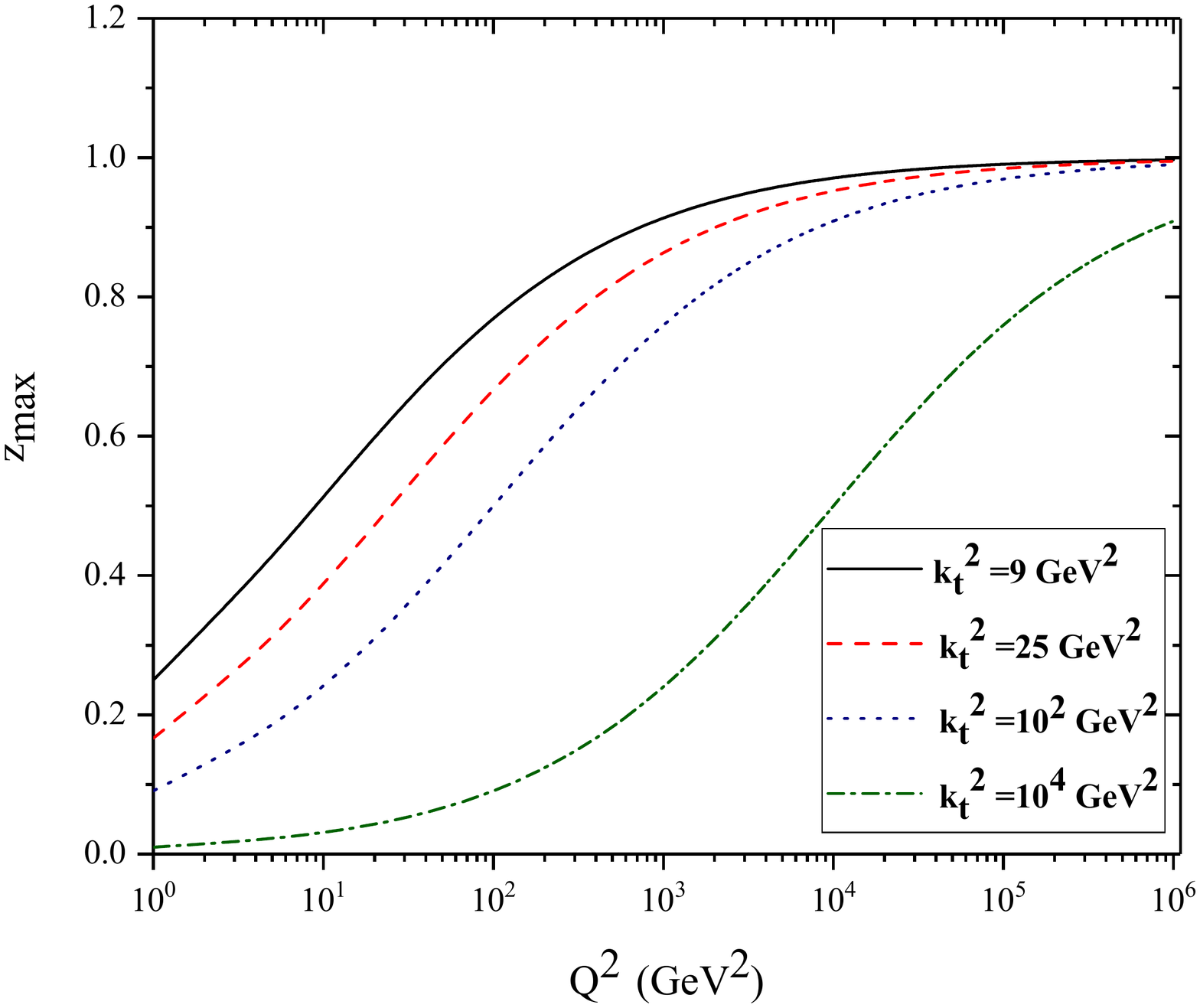}
}
\caption{The $z_{max}$ versus the hard scale $Q^2$ \cite{Modarres1}. See the text for more explanations.
}\label{zmax}
\end{center}
\end{figure}
For all $x$ values  a good agreement is observed between the  different groups input PDFs sets, i.e.,  UPDFs, especially at large scale, which is rooted in the behavior of PDFs sets in the large hard scales. We should mention here that the Sudakov form factors are independent of the input PDFs sets, therefore, in the UPDFs ratios they are canceled out. On the other hand,
the parton distributions of  quarks and gluon in the equations (\ref{fq}) and (\ref{fg}) (in the form $xq(\frac{x}{z}, k_t^2)$, $x\bar{q}(\frac{x}{z}, k_t^2)$  and $xg(\frac{x}{z}, k_t^2)$) enter the calculations at the scale $\mu^2 =k_t^2$. This causes the divergent behavior of different PDFs sets in the small scale \citep{Modarres1} which affect the functional forms of UPDFs in the small $k_t^2$. Therefore, the result is an increase in the disagreement of the UPDFs with a decrease in  $k_t^2$. In addition, the steeper and divergent in the UPDFs ratios  appear in the larger $x$ values, which show themselves with increasing $k_t^2$ and move to larger $Q^2$. This divergence is rooted in the definition of $z_{max}$ (see the figure
 \ref{zmax} \cite{Modarres1}). In fact, this divergence occurs in the regions where the upper and lower limits of the integral are close to each other ($x \simeq z_{max}$) and depending on each UDPFs, tends to zero rapidly. So the ratio of UPDFs becomes zero or infinite. According to definition of $z_{max}=1-\Delta ={Q \over (Q+k_t )}$, the value of $Q$ around which this divergence occurs can be obtained for fixed $x$ and $k_t$, which is similar  to the results of reference \citep{Modarres1}. 
 
One can observe from  the panels (e) and (f) of the figure \ref{updf1} that the ratio of PB TMDPDFs to the corresponding UPDFs using MMHT2014 PDFs set, has different behaviors   with respect to the other ratios in the panels of figures   \ref{updf0} and \ref{updf1}. This shows that the PB TMDPDFs are calculated with different and stronger constraint and cut-off with respect to that of KMR formalism. For example, since  PB TMDPDFs become approximately zero for $Q^2< k_t^2$, the ratios go to zero in this range, but for  $Q^2>k_t^2$ they become larger with respect 
to other panels ratios. This behavior roots to the normalization condition of the equation (\ref{nor})
(see the comments above this equation about the $k_t^2$ factor) and different constraint which imposed on the PB TMDPDFs. On the other hand, it uses HERAPDF20 PDFs which behave differently regarding the other PDFs sets.

\subsection{The Gaussian transverse momentum dependent fit to the KMR UPDFS }\label{subsection3.2}
As we pointed out in the introduction, the $k_t^2$ dependence of UPDFs, both in the perturbative and non-perturbative regions are usually assumed to have Gaussian shape \cite{PBTMD,Metz,Bacchetta,T1,T2} with Gaussian width around $<k_t^2>= 0.25-0.44$ $GeV^2$, at the hard scale equal to $2.4$ $GeV^2$. So it would be interesting to find out if the  KMR UPDFs to have similar behavior.
To calculate the average transverse momentum $<k_t^2>$, we plot the UPDFs using the KMR method for a constant scale $Q^2$ and    $ 0.0045<x <0.1$, by using   MSHT20 PDFs set in the LO approximation. 
The functional form that we  fit to the UPDFs is as following (note that we multiply the Gaussian function by $k_t^2$ and omit the $\pi$ in the denominator of the equation (\ref{q1}), because of  our definitions of UPDFs and the integration, i.e., $dk_t^2$ rather than $d^2k_t$) :
\begin{equation}
f_a(x,k_t^2,\mu^2)=f_a(x,\mu^2){k_t^2 \over <k_t^2>} exp\left(-{k_t^2 \over <k_t^2>}\right).\label{G2}
\end{equation}
Then the normalization condition in the KMR method, i.e.,:
\begin{equation}
\int_{0}^{\mu^2}{dk_t^2 \over k_t^2} f_a(x,k_t^2,\mu^2) =f_a(x,\mu^2),\label{nor}
\end{equation}
is easily satisfied (where $f_a(x,\mu^2)$ are the different PDFs sets). So, the Gaussian widths $<k_t^2>$ can be obtained for different $x$ values by fitting and  averaging between them,  are presented as the average transverse momentum $<k_t^2>$ at  each scale $Q^2$. 
\begin{table}[h!]
  \begin{center}
    \caption{The values of average transverse momentum for the light quarks and gluon in the hard scales $Q^2=2.4$ to $1.44\times10^4$ $GeV^2$.}
    \label{tab:table1}
    \begin{tabular}{|l|l|l|l|l|}
    \hline
    \textbf{$ Q^2\; (GeV^2)$} & \textbf{$ <k_t^2>_u\; (GeV^2)$} & \textbf{$<k_t^2>_d\; (GeV^2)$} & \textbf{$<k_t^2>_s\; (GeV^2)$}& \textbf{$<k_t^2>_g\; (GeV^2)$}\\
    \hline
    2.4 & 0.33 & 0.32 & 0.32 & 0.41\\
    \hline
    9 & 0.40 & 0.38 & 0.39 & 0.5\\
    \hline
    25 & 0.46 & 0.44 & 0.45 & 0.57\\
    \hline
    49 & 0.5 & 0.48 & 0.48 & 0.62\\
    \hline
    81 & 0.53 & 0.50 & 0.51 & 0.66\\
    \hline
    100 & 0.54 & 0.51 & 0.53 & 0.68\\
    \hline
    400 & 0.63 & 0.60 & 0.61 & 0.79\\
    \hline
    900 & 0.69 & 0.65 & 0.67 & 0.86\\
    \hline
    1600 & 0.73 & 0.7 & 0.72 & 0.92\\
    \hline
    2500 & 0.76 & 0.73 & 0.75 & 0.97\\
    \hline
    3600 & 0.79 & 0.76 & 0.78 & 1.01\\
    \hline
    4900 & 0.82 & 0.79 & 0.81 & 1.05\\
    \hline
   6400 & 0.84 & 0.81 & 0.83 & 1.10\\
   \hline
    8100 & 0.87 & 0.83 & 0.85 & 1.12\\
    \hline
    10000 & 0.89 & 0.85 & 0.87 & 1.15\\
    \hline
    12100 & 0.90 & 0.87 & 0.89 & 1.18\\
    \hline
    14400 & 0.92 & 0.88 & 0.91 & 1.21\\
    \hline
     \end{tabular}
  \end{center}
\end{table}
\begin{figure}
\begin{center}
\includegraphics[height =8cm]{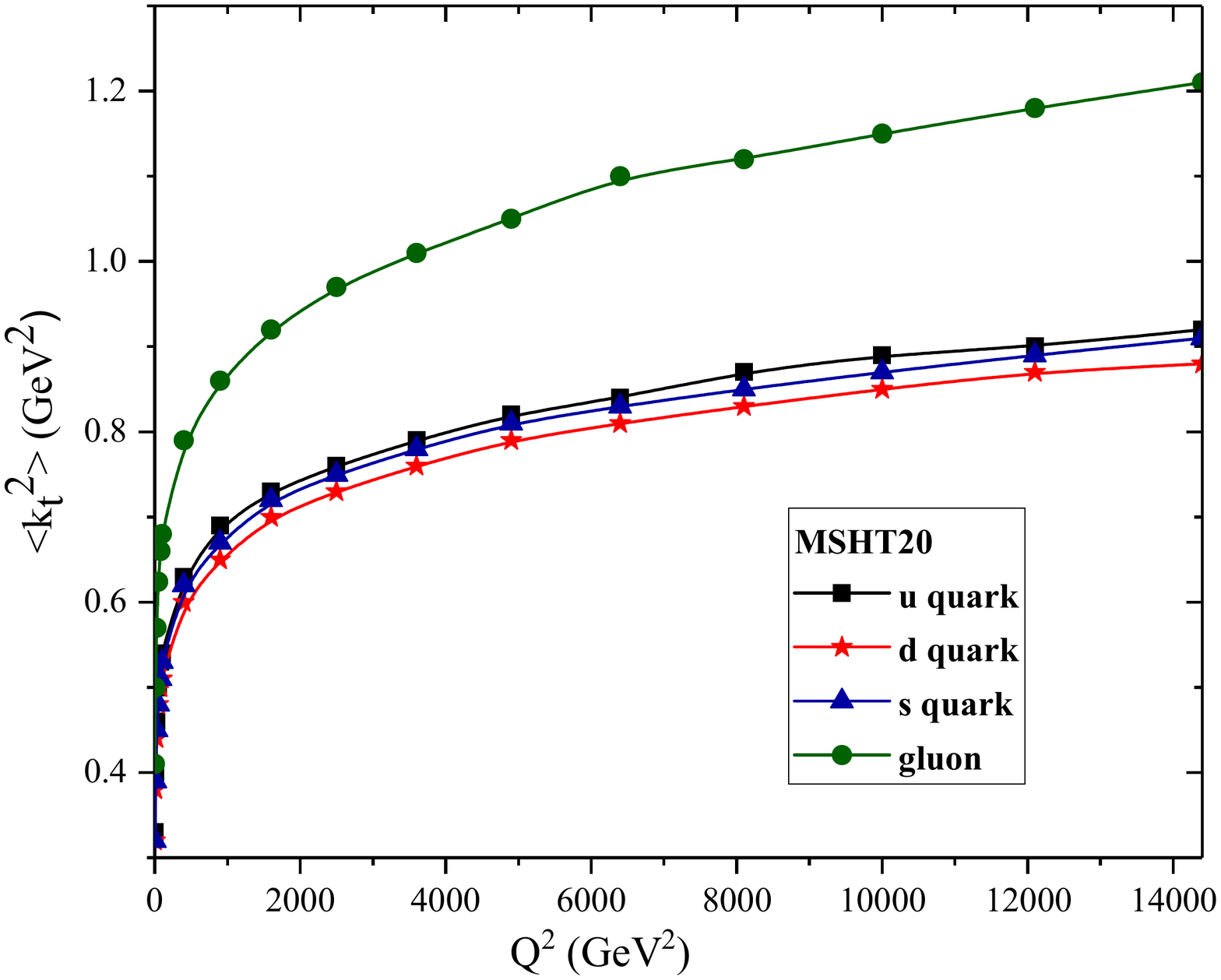}
\caption{The average transverse momentum  $<k_t^2>$ in the range $Q^2 =2.4$ to $1.44\times10^4$ $GeV^2$.}
\label{fig:2}
\end{center}
\end{figure}
We perform the above calculations for   light quarks and gluon in the  range $Q^2=2.4$ to $1.44\times10^4$ $GeV^2$ and the results are given in the table \ref{tab:table1} and plotted in the figure \ref{fig:2}. It is seen that for the light quarks the variation of $<k_t^2>$ with respect to $Q^2$ is approximately the same, while for gluon it becomes larger. This can be resulted from the general behavior of the gluon PDFs with respect  to quarks ones.

\begin{figure}
\begin{center}
\subfigure[]{
\includegraphics[height =6cm]{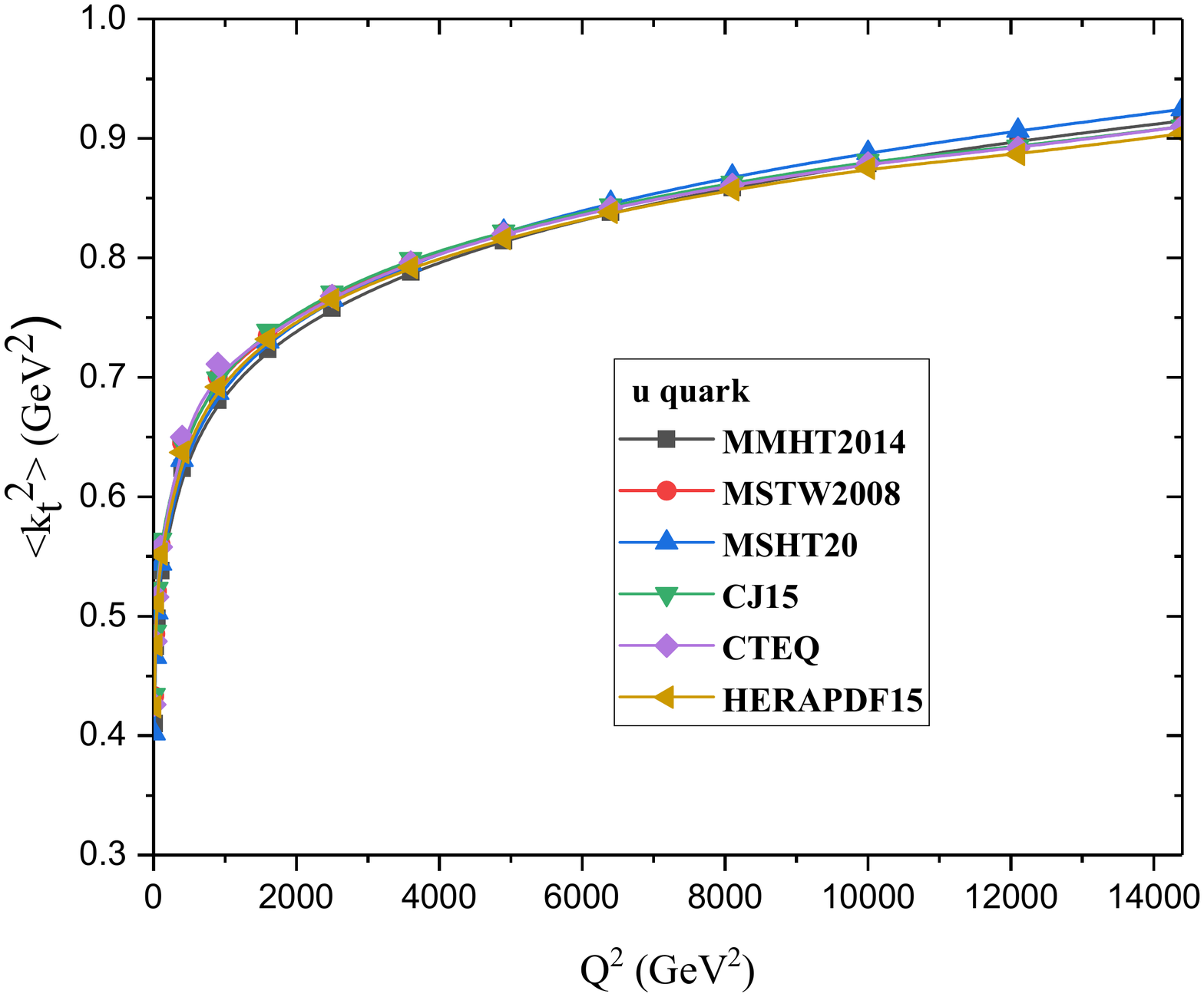}
}
\subfigure[]{
\includegraphics[height =6cm]{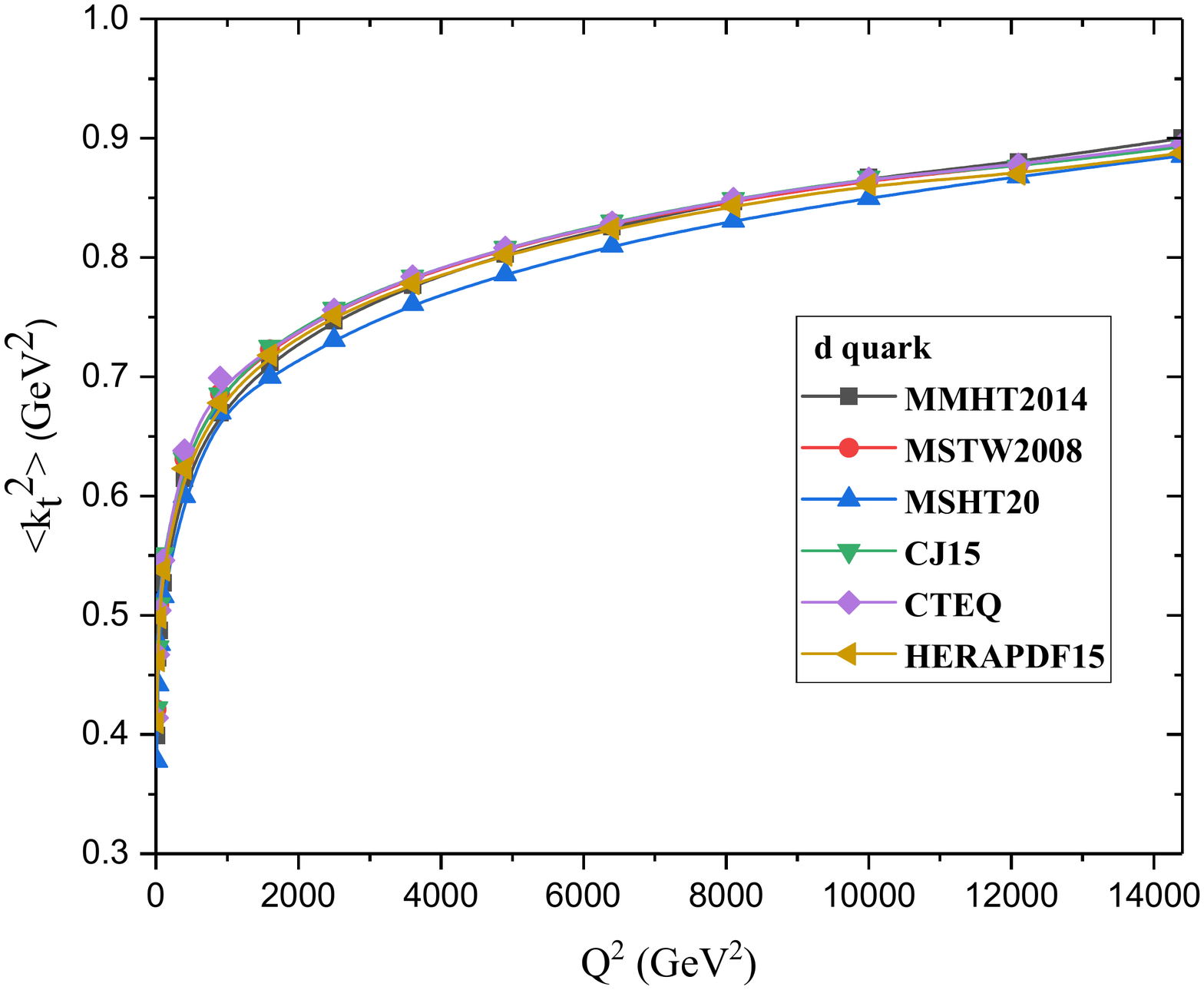}
}
\subfigure[]{
\includegraphics[height =6cm]{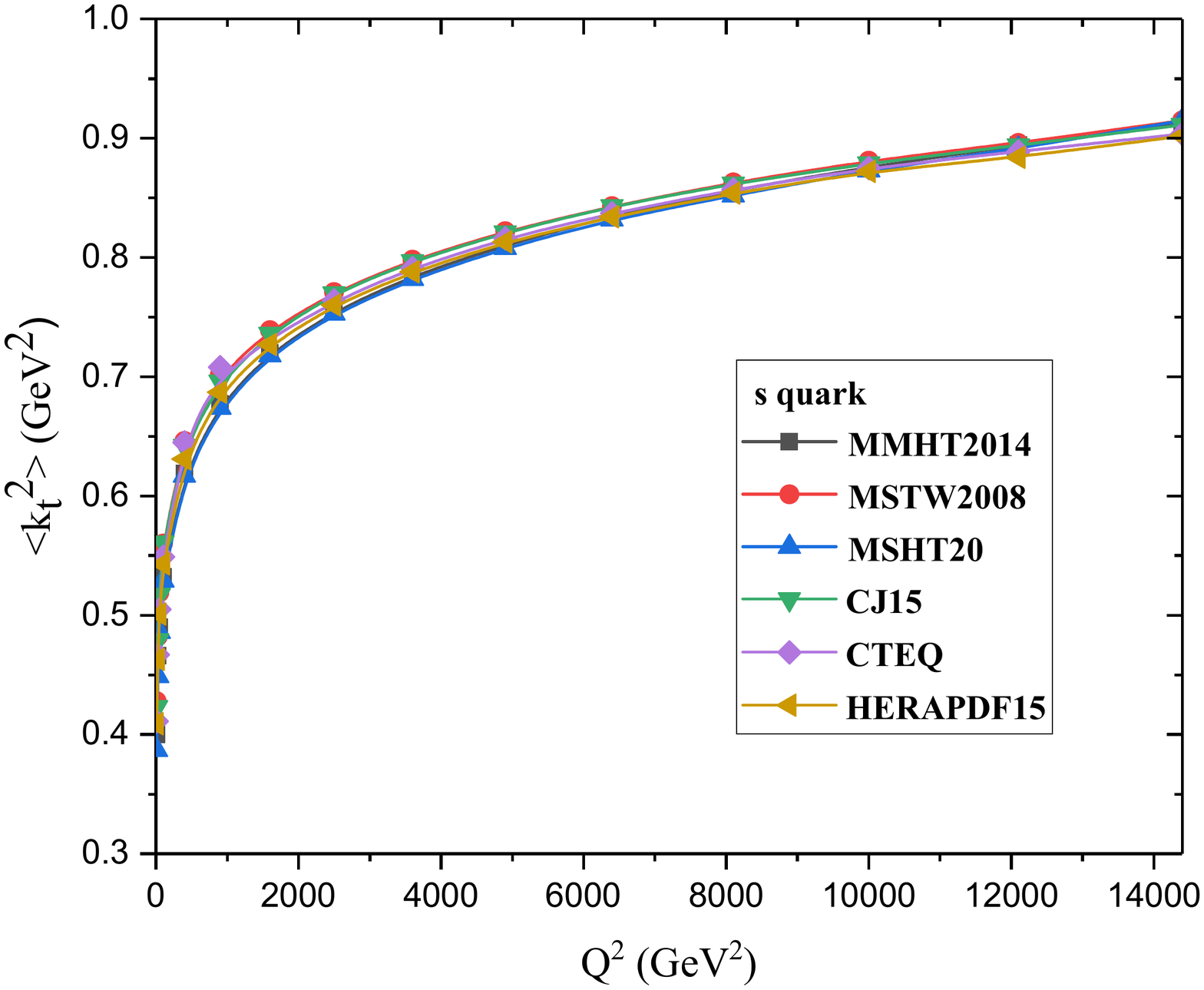}
}
\caption{The plots of $<k_t^2>$ for the light quarks, using different PDFs sets: (a) u quark, (b) d quark, (c) s quark.}
\label{fig:3}
\end{center}
\end{figure}

\begin{figure}
\begin{center}
\includegraphics[height =8cm]{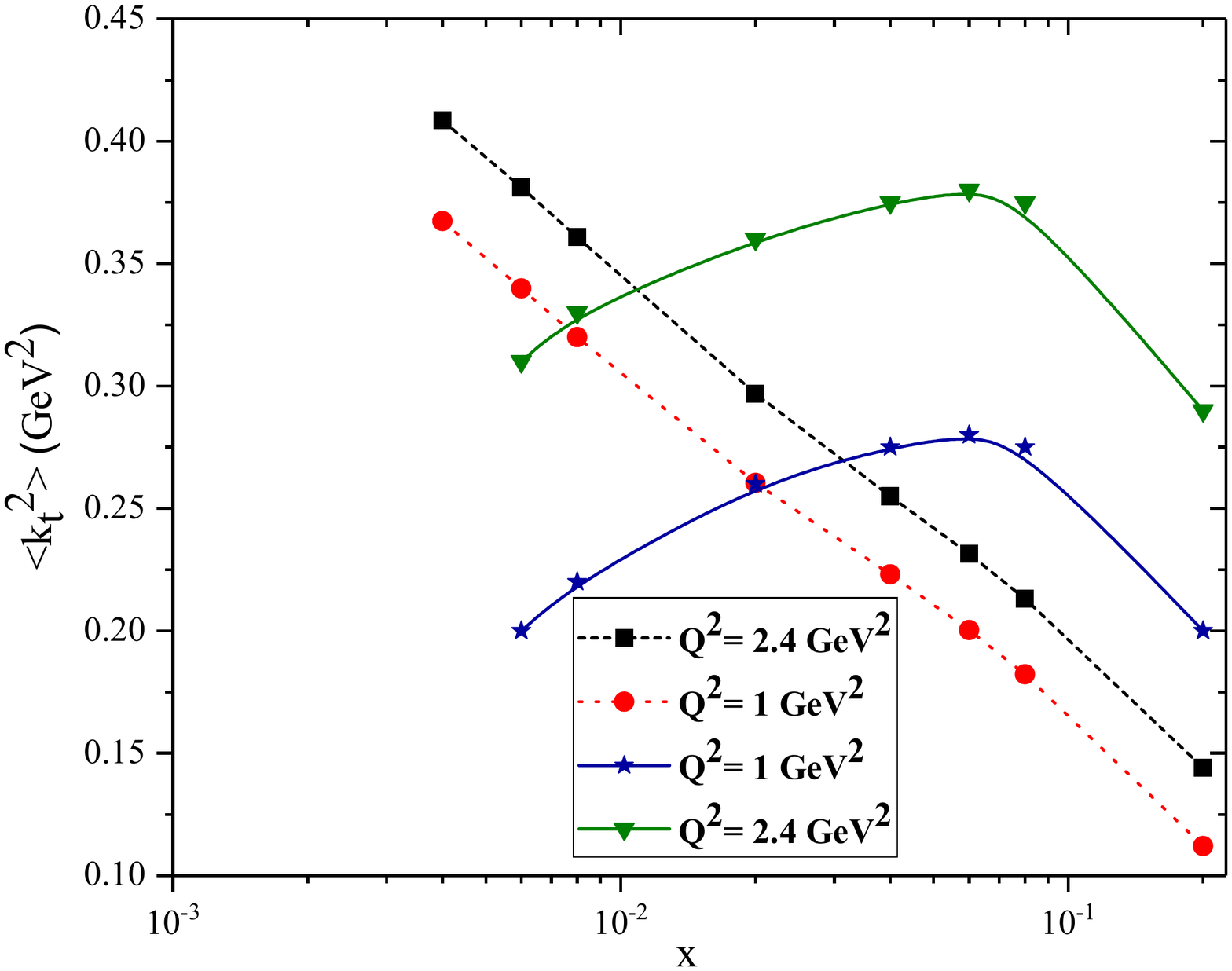}
\caption{The plot of $<k_t^2>$ versus $x$ for the average transverse momentum of u quark   (dotted curves) and those of reference \citep{Bacchetta} (full curves). }
\label{fig:4}
\end{center}
\end{figure}

Then, we repeat the above calculation (fitting the UPDFs to the equation (\ref{G2}))  for MMHT2014, CTEQ6l1, CJ15, HERAPDF15 and MSTW2008 PDFs sets   selected from LHAPDF6 library \cite{LHAPDF}. The results of these   average transverse momenta for the light quarks are also shown in the figure \ref{fig:3} to see if the resulted average transverse momentum      depends on the choice of different input PDFs sets. The similar behaviors with respect to various PDFs sets are found for u, d and s quarks.
In addition  we also explore the $x$ dependence of the average transverse momentum \citep{Bacchetta} which are presented in the figure \ref{fig:4}. There is a reasonable agreement between our result and those of reference \citep{Bacchetta}, especially in the higher $x$ values, considering the size of $<k_t^2>$ and this fact that we are not using any free  parameter in our UPDFs beside the input PDFs sets.
\subsection{The proton structure function in the $k_t$-factorization framework}\label{subsection3.3}
\begin{figure}
\begin{center}
\subfigure[]{
\includegraphics[height =6cm]{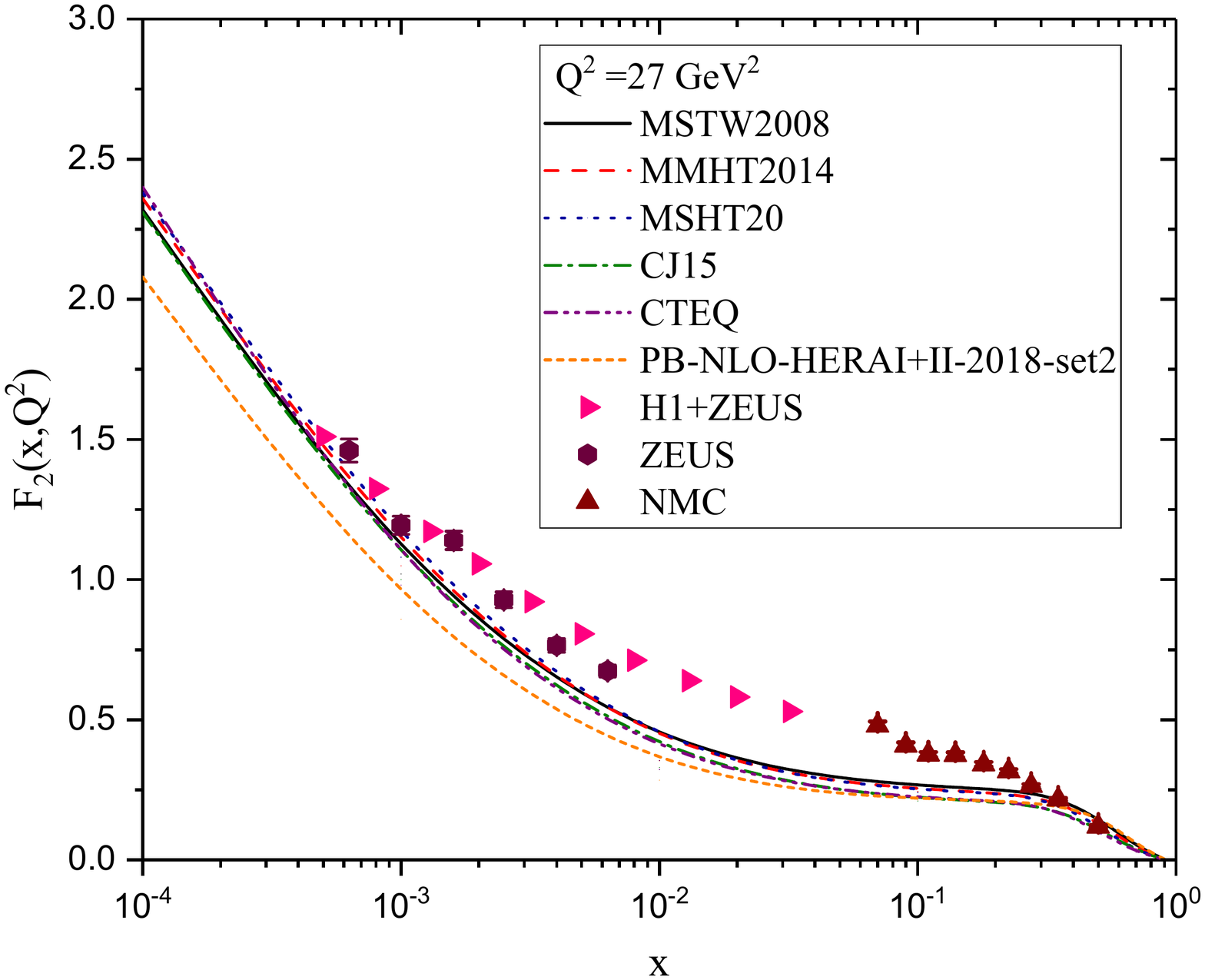}
}
\subfigure[]{
\includegraphics[height =6cm]{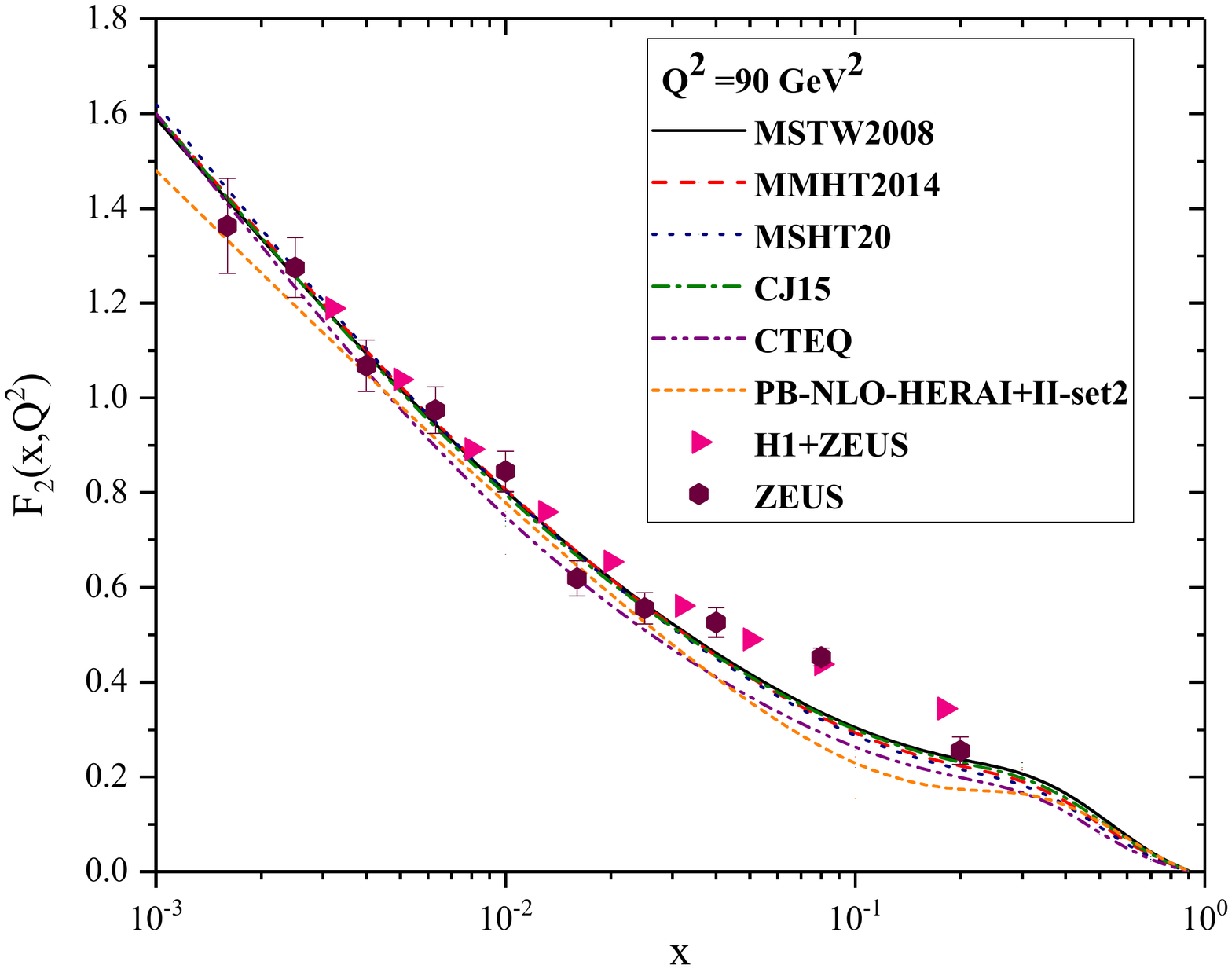}
}
\caption{Compare the structure function for numerous sets of PDFs with experimental data in the scale (a)  $Q^2 =27$ $GeV^2$ and (b) $Q^2 =90$ $GeV^2$.}
\label{fig:6}
\end{center}
\end{figure}

For more detailed and precise  studies, it is necessary to examine how the different UPDFs related to  each PDFs sets, behave  in the $k_t$-factorization  phenomenological calculations in comparison to a corresponding measured observable, i.e.,    the $ep$ SIDIS experimental data. In recent years, the proton structure function can be measured with good accuracy  up to the small $x$ values and the very high hard  scale.   So,  we calculate and plot the  proton structure functions, using   the KMR UPDFs in the $k_t$-factorization framework with the different  PDFs sets such as  MSHT20, MMHT2014, MSTW2008, CJ15 and CTEQ6l1,   for the two scales $Q^2 =27$ and  $90$ $GeV^2$      and compare the results   with the experimental data  of the NMC, ZEUS and H1+ZEUS collaborations in the figure \ref{fig:6}.  As one should expect there is good  agreement between our predictions and the data at higher hard scale value, i.e., $90$ $GeV^2$. On the other hand there are not much difference between the various input PDFs sets (we do not consider other PDFs sets because of computer time consuming). The SF with PB TMDPDF (PB-NLO-HERAI+II-set2) \cite{PBTMD} is also evaluated and plotted.  

\begin{figure}
\begin{center}
\subfigure[]{
\includegraphics[height =6cm]{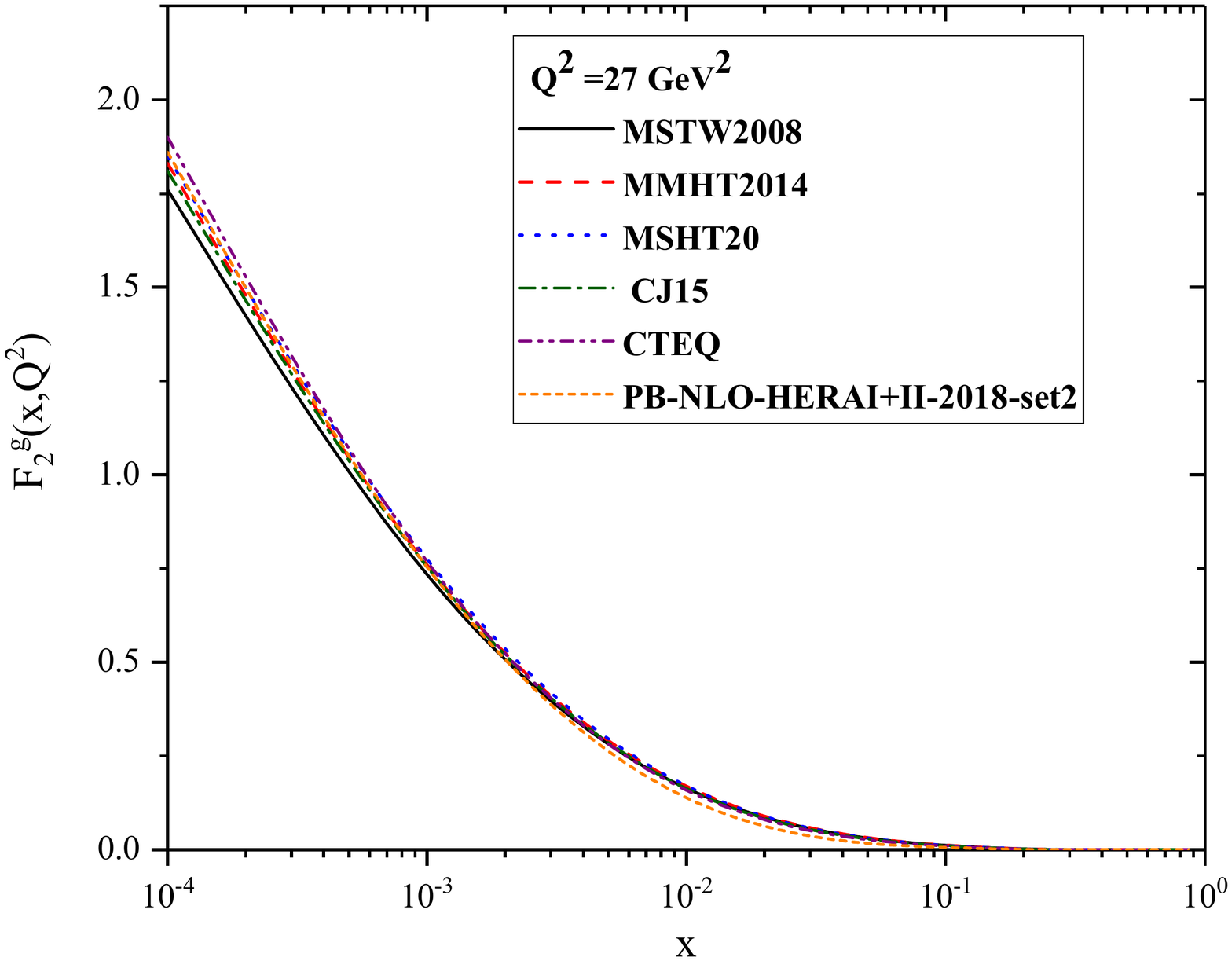}
}
\subfigure[]{
\includegraphics[height =6cm]{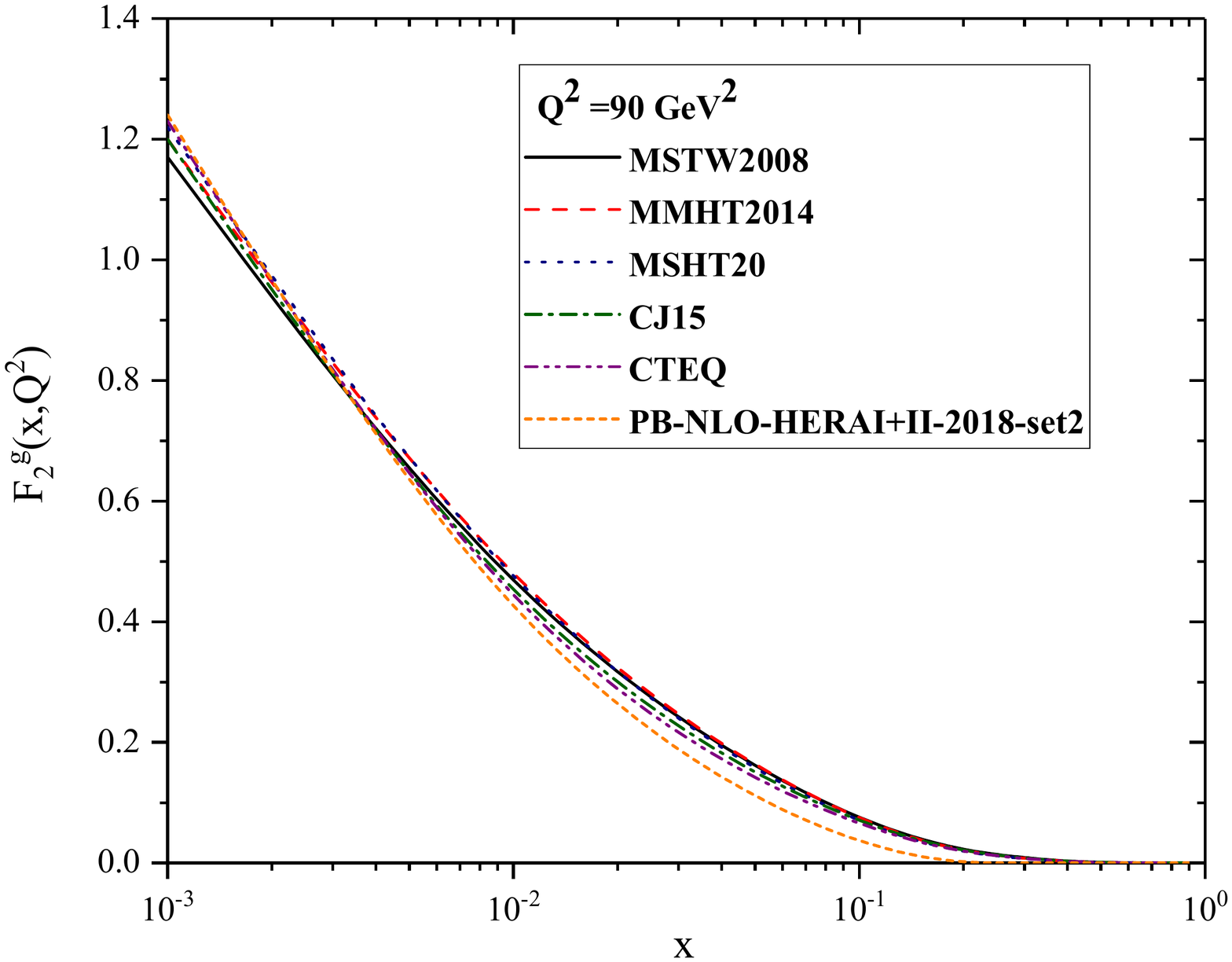}
}
\caption{The gluon contribution to structure function of proton for numerous sets of PDFs in the scale (a)   $Q^2 =27$ $GeV^2$ and (b) $Q^2 =90$ $GeV^2$. }
\label{fig:5}
\end{center}
\end{figure}

\begin{figure}
\begin{center}
\subfigure[]{
\includegraphics[height =6cm]{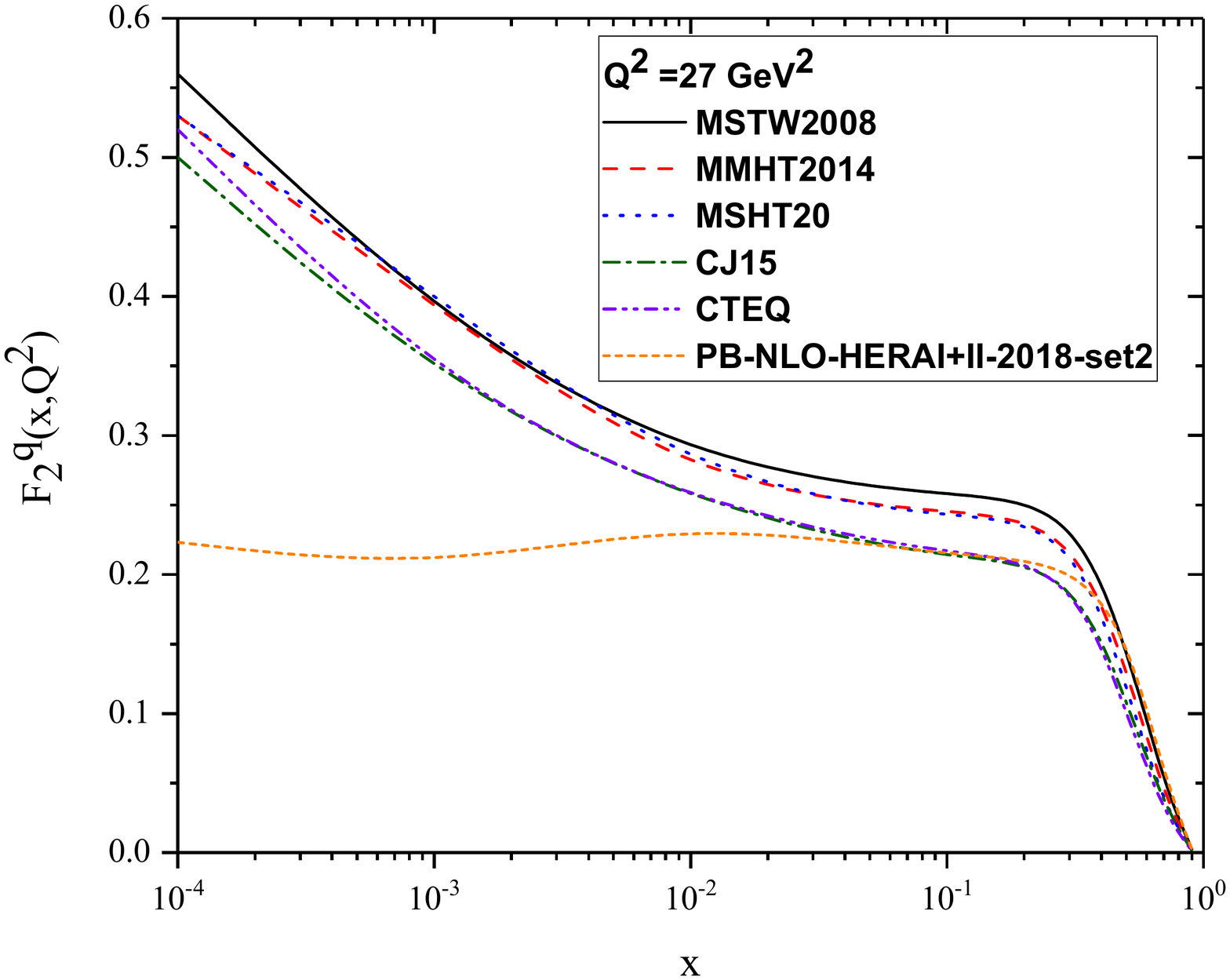}
}
\subfigure[]{
\includegraphics[height =6cm]{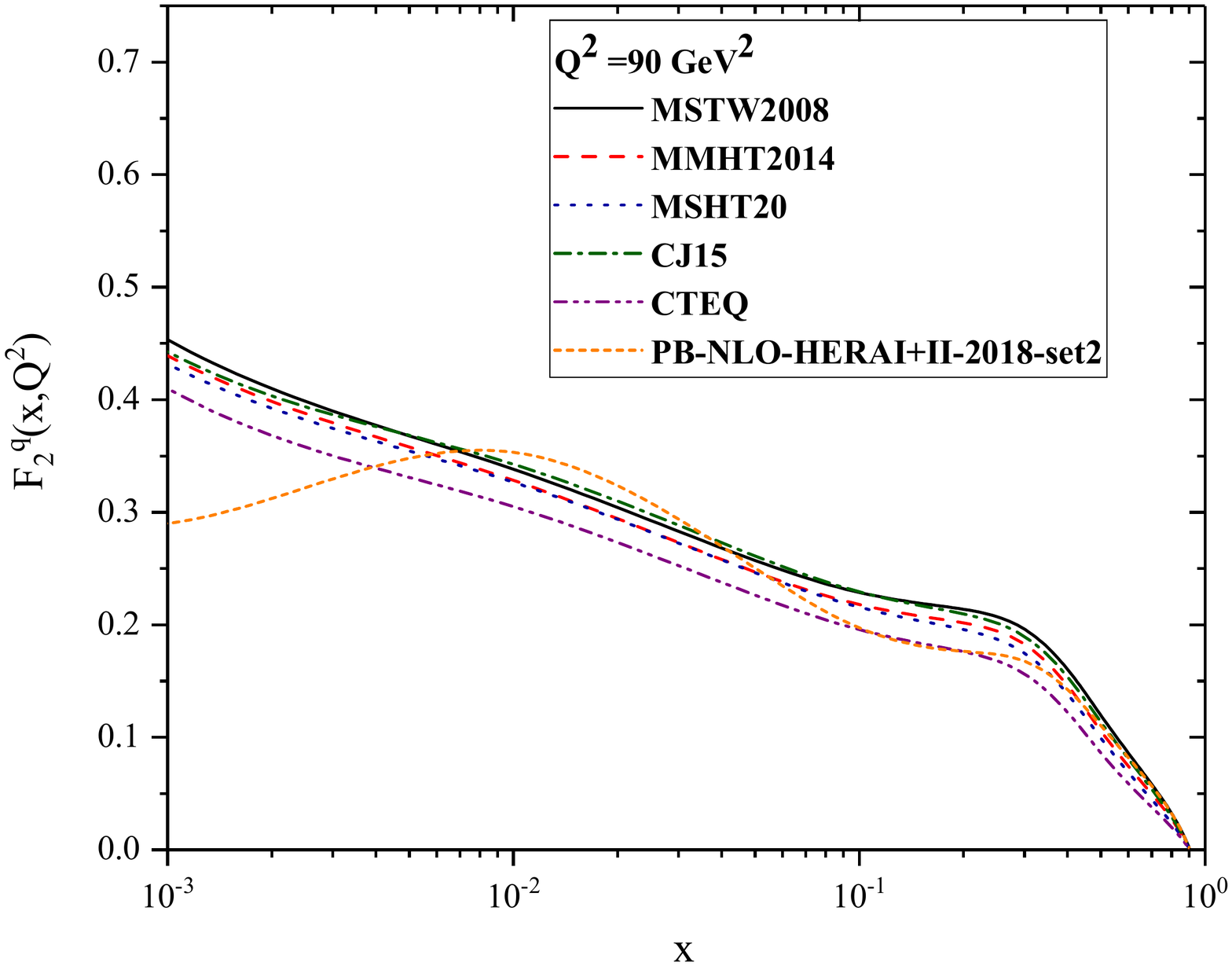}
}
\caption{The quark contribution to structure function of proton for numerous sets of PDF in the scale (a) $Q^2 =27$ $GeV^2$ and (b) $Q^2 =90$ $GeV^2$. }
\label{fig:5p}
\end{center}
\end{figure}

The gluon and quarks contribution to the SF of proton are plotted in the figures \ref{fig:5} and \ref{fig:5p} to show their behaviors in terms of various PDFs sets. As one should expect , it is observed that the main contribution is comping from the gluon in the small $x$ region and again there are not much differences between the various PDFs sets. While the calculation with  PB TMDPDFs give the smaller quarks distribution and larger gluon ones. This different behaviors are due to PB TMDPDFs which was discussed before in connection to the figure 3 and as it is claimed  a global fitting is performed.

\begin{figure}
\begin{center}
\subfigure[]{
\includegraphics[height =6cm]{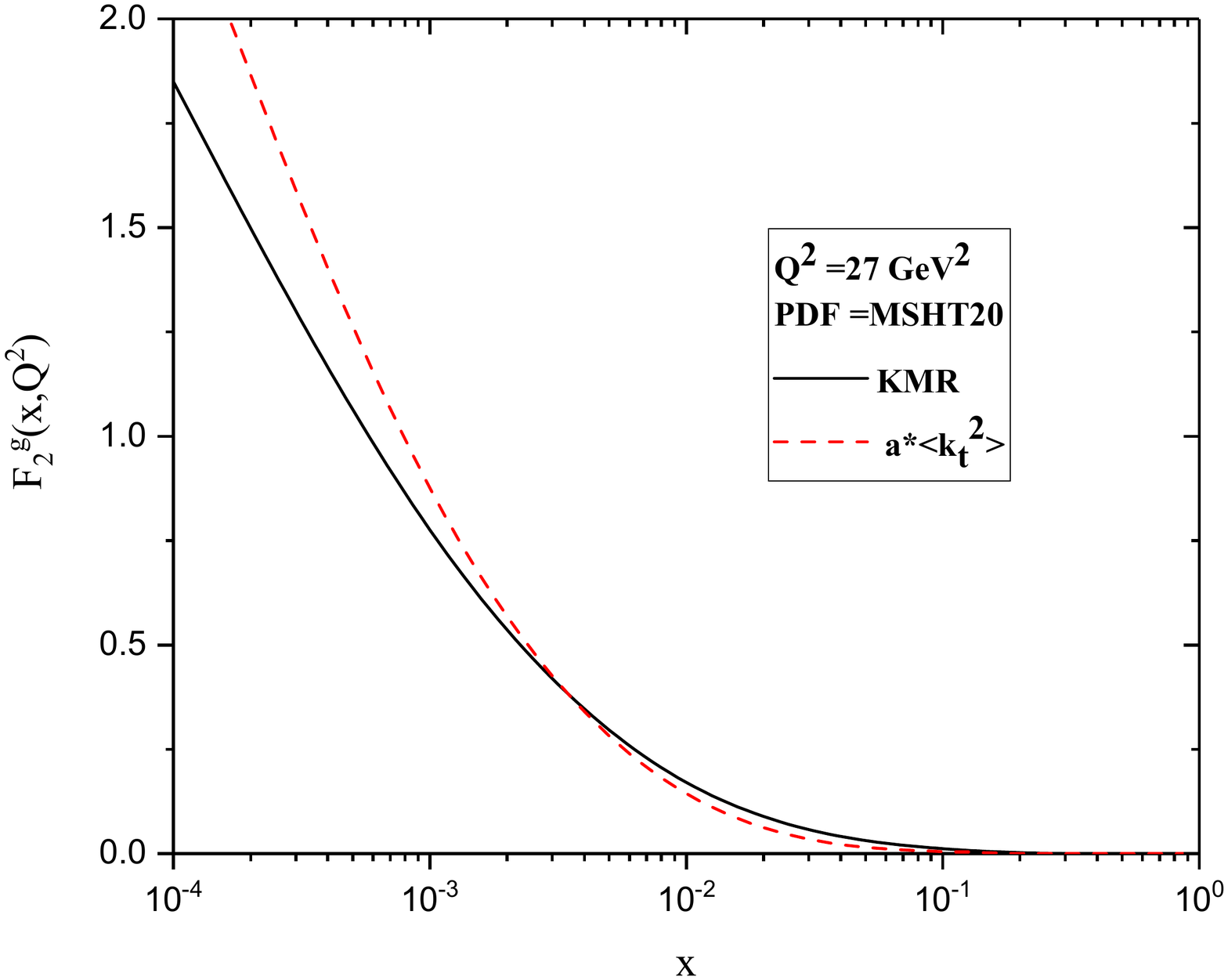}
}
\subfigure[]{
\includegraphics[height =6cm]{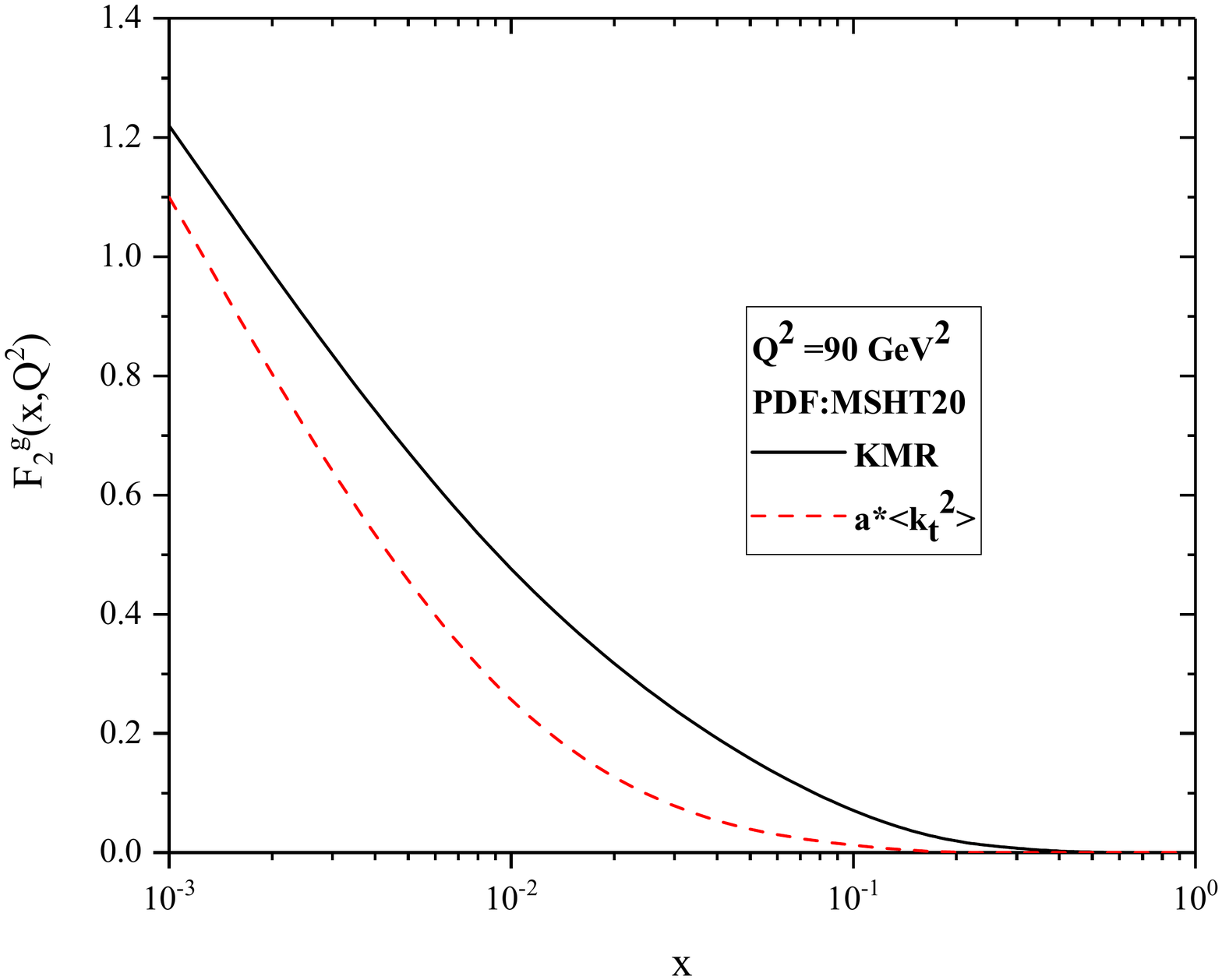}
}
\caption{The comparison of gluon contribution to the structure function of proton, using KMR-UPDFs and Gaussian dependent transverse momentum (discussed in the section \ref{subsection3.2}) multiply by the corresponding PDFs set in the hard scale a) $Q^2 =27$ $GeV^2$ and b) $Q^2 =90$ $GeV^2$ with $a=39.5$. }
\label{fig:7}
\end{center}
\end{figure}

Finally, we obtain and plot, the gluon contribution to the proton structure function   using the Gaussian  transverse momentum dependent, discussed in the section \ref{subsection3.2} multiplied by corresponding PDFs sets in comparison with the KMR UPDFs input, see the figure  \ref{fig:7}. The average traverse momentum $<k_t^2>$ is used for the corresponding $x$ values and hard scale $Q^2$. We multiply $<k_t^2>$ by a constant, i.e., $a=39.5$ to get the best fit to KMR UPDFs SF at $Q^2=27$ $GeV^2$.  It is seen that a reasonably result is achieved with the Gaussian assumption for the transverse dependent behavior of the UPDFs. Note that we use the same $a=39.5$ for the panel (b).
\section{\textbf{Conclusions}}
In this work, we obtained the ratio of KMR UPDFs for the several sets of PDFs  in the leading order with respect to the KMR UPDFs considering MMHT2014 PDFs set. It was shown that    the KMR UPDFs with different input PDFs sets  are almost  very similar and stable  \citep{Modarres1}. However the PB TMDPDFs have    different behavior with respect to the KMR UPDFs, both by changing the hard scale and the transverse momentum.

In addition for the first time, we obtained the average transverse momentum in the high energy scales by fitting the $k_t$-dependent of the  KMR UPDFs at each $x$ and $Q^2$ . It was shown that the average transverse momentum is not very sensitive to the longitudinal momentum fraction $x$, so one can average them with respect to $x$ at each scale $Q^2$. On the other hand, it   does not have much variation regarding  to the various quark flavours, i.e., they are almost the same for the light quarks, which is consistent to the results of references \citep{Metz,Signori} (the section \ref{subsection3.2}). Its sensitivity considering   the different input PDFs  sets   was also tested.
The  value of the average transverse momentum in the scale $ Q^2 =2.4$ $GeV^2 $ is exactly within the range reported in the  reference \citep{Metz}. However, it  is greater than the one was fitted and used in the parton branching method, i.e., $0.125$ $GeV^2$ \cite{PBTMD}. 

In the  section \ref{subsection3.3}, for a detailed examination, we calculated the proton structure function for a number of PDFs sets by using the KMR UPDFs and the $k_t$-factorization formalism  and compared them with the experimental data. The results showed that the structure function calculated by KMR method is  consistent with the experimental results and it is reasonably independent of the different PDFs sets used as the input.  While the quarks and gluon contribution to proton SF using PB TMDPDFs, behave differently. Also we obtained the gluon contribution of the structure function using the KMR method and Gaussian transverse momentum dependent multiplied with corresponding PDFs sets, which showed  a proper agreement using a constant variation for the best fitting.

We should make this remark that only the LO level analysis was performed in this report since it was intended to show the relative behaviors of different PDFs and UPDFs set. However the whole calculations can be extended to the NLO level. Therefore, in our future works, by calculating  the NLO level contributions to the SF, we could have a better conclusion about the disagreement in the middle $x$ region of proton SF.
\appendix
\section{J values}
\label{a}
\begin{equation}\tag{a-1}
J_1 =\frac{\kappa_t^2}{D_1^2}
\end{equation}
\begin{equation}\tag{a-2}
J_2 =\frac{1}{\sqrt{(D_1+k_t^2)^2-4\kappa_t^2 k_t^2}}-\frac{(D1+k_t^2)(\beta(1-\beta)Q^2+m_q^2)}{[(D_1+k_t^2)^2-4\kappa_t^2 k_t^2]^{3/2}}
\end{equation}
\begin{equation}\tag{a-3}
J_3 =\frac{1}{D_1}\left( 1+\frac{\kappa_t^2-\beta(1-\beta)Q^2-m_q^2-k_t^2}{\sqrt{(D_1+k_t^2)^2-4\kappa_t^2 k_t^2}}\right) 
\end{equation}
\begin{equation}\tag{a-4}
J_4 =\frac{1}{D_1^2}
\end{equation}
\begin{equation}\tag{a-5}
J_5 =\frac{D_1+k_t^2}{[(D_1+k_t^2)^2-4\kappa_t^2 k_t^2]^{3/2}}
\end{equation}
\begin{equation}\tag{a-6}
J_6 =\frac{2}{D_1} \frac{1}{\sqrt{(D_1+k_t^2)^2-4\kappa_t^2 k_t^2}}
\end{equation}
That 
\begin{equation}\tag{a-7}
D_1 =\kappa_t^2+ \beta(1-\beta)Q^2+m_q^2
\end{equation}

 
\newpage
\end{document}